\documentstyle[10pt,epsf,epsfig,hangcaption,xspace,amssymb,amsfonts,amsmath,amsthm,cite,
dp_delphititle]{dp_delphi}
\makeindex
\def\DpPaperGroup{PH-EP}
\def\DpPaperRef{2005-020}
\def\DpDate{27 April 2005}
\def\DpAuthors{DELPHI Collaboration}
\def\DpSubmit{(Accepted by Euro. Phys. Journ. C)}
\def\DpTitle{{Determination of the $b$ quark mass at the $M_Z$ scale with the DELPHI detector at LEP }}
\def\DpComment{ }
\def\DpEMail{ }
\newfont{\scsl}{ecsc1200} % scaled 1000
\newcommand{\Zfitter}{{\scsl{%
\raisebox{-0.4ex}{z\kern-0.05em{}f}\kern-0.1em{}I\kern-0.15em%
\raisebox{0.8ex}{T}\kern-0.25em{}T\kern-0.25em%
\raisebox{-0.8ex}{E\kern-0.05em{}r}}}}
\newcommand{\permil}{\raisebox{0.5ex}{\tiny $0$}$\!/$\raisebox{-0.3ex}{\tiny $\! 00$}}

\def\ZP{Z.\ Phys.\ {\bf C}}
\def\PL{Phys.\ Lett.\ {\bf B}}
\def\PR{Phys.\ Rev.\ {\bf D}}

\def\NP{Nucl.\ Phys.\ {\bf B}}

\def\NIM{Nucl.\ Instr.\ and Meth.\ {\bf A}}
\def\be{\begin{equation}}
\def\ee{\end{equation}}
\def\bea{\begin{eqnarray}}
\def\eea{\end{eqnarray}}

\begin{document}
%%%%%%%%%%%%%%%%%%%%%%%%%% They are a problem with Coll.Sty ?
\makeatletter
%\input{dp_system:coll.sty}
% Collapse citation numbers to ranges.  Non-numeric and undefined labels
% are handled.  No sorting is done.  E.g., 1,3,2,3,4,5,foo,1,2,3,?,4,5
% gives 1,3,2-5,foo,1-3,?,4,5
\newcount\@tempcntc
\def\@citex[#1]#2{\if@filesw\immediate\write\@auxout{\string\citation{#2}}\fi
  \@tempcnta\z@\@tempcntb\m@ne\def\@citea{}\@cite{\@for\@citeb:=#2\do
    {\@ifundefined
       {b@\@citeb}{\@citeo\@tempcntb\m@ne\@citea\def\@citea{,}{\bf ?}\@warning
       {Citation `\@citeb' on page \thepage \space undefined}}%
    {\setbox\z@\hbox{\global\@tempcntc0\csname b@\@citeb\endcsname\relax}%
     \ifnum\@tempcntc=\z@ \@citeo\@tempcntb\m@ne
       \@citea\def\@citea{,}\hbox{\csname b@\@citeb\endcsname}%
     \else
      \advance\@tempcntb\@ne
      \ifnum\@tempcntb=\@tempcntc
      \else\advance\@tempcntb\m@ne\@citeo
      \@tempcnta\@tempcntc\@tempcntb\@tempcntc\fi\fi}}\@citeo}{#1}}
\def\@citeo{\ifnum\@tempcnta>\@tempcntb\else\@citea\def\@citea{,}%
  \ifnum\@tempcnta=\@tempcntb\the\@tempcnta\else
   {\advance\@tempcnta\@ne\ifnum\@tempcnta=\@tempcntb \else \def\@citea{--}\fi
    \advance\@tempcnta\m@ne\the\@tempcnta\@citea\the\@tempcntb}\fi\fi}
 
\makeatother
%%%%%%%%%%%%%%%%%%%%%%%%%% ??????????????????????????????????
% Generate the title page
\begin{titlepage}
\pagenumbering{roman}
\CERNpreprint{\DpPaperGroup}{\DpPaperRef} % Reference of the paper
\date{{\small\DpDate}} % Date of the paper
\title{\DpTitle} % Title of the paper
\address{\DpAuthors} % General name of the author(s)
\begin{shortabs} % Start the abstract
\noindent
An experimental study of the normalized three-jet rate of $b$ quark
events with respect to light quarks events (light= $\ell \equiv
u,d,s$) has been performed using the {\sc Cambridge} and {\sc Durham}
jet algorithms. The data used were collected by the {\sc Delphi}
experiment at LEP on the $Z$ peak from 1994 to 2000. The results are
found to agree with theoretical predictions treating mass corrections
at next-to-leading order. Measurements of the $b$ quark mass have also
been performed for both the $b$ pole mass: $M_b$ and the $b$ running
mass: $m_b(M_Z)$. Data are found to be better described when using the
running mass. The measurement yields:

\[
m_b(M_Z) = 
2.85 \pm 0.18 ~({\rm stat}) \pm 0.13 ~({\rm exp})
\pm 0.19 ~({\rm had}) \pm 0.12 ~({\rm theo})~{\rm GeV}/c^2
\] 
for the {\sc Cambridge} algorithm.
%where (stat), (exp), (had) and (theo) refer to the estimated uncertainties due to the
%limited statistics in the event sample, to the experimental and detector correction procedure,
%to the hadronization process and to the precision of the theoretical calculations, respectively.

This result is the most precise measurement of the $b$ mass derived
from a high energy process. When compared to other $b$ mass
determinations by experiments at lower energy scales, this value
agrees with the prediction of Quantum Chromodynamics for the energy
evolution of the running mass. The mass measurement is equivalent to a
test of the flavour independence of the strong coupling constant with
an accuracy of 7\permil.

\end{shortabs}
\vfill
\begin{center}
\DpSubmit \ \\ % Horrible hack to allow to have DpSubmit empty
\DpComment \ \\
\DpEMail \ \\
\end{center}
\vfill
\clearpage
\headsep 10.0pt
\addtolength{\textheight}{10mm}
\addtolength{\footskip}{-5mm}
\begingroup
% Commands to process the author names
%
\newcommand{\DpName}[2]{\hbox{#1$^{\ref{#2}}$},\hfill}
\newcommand{\DpNameTwo}[3]{\hbox{#1$^{\ref{#2},\ref{#3}}$},\hfill}
\newcommand{\DpNameThree}[4]{\hbox{#1$^{\ref{#2},\ref{#3},\ref{#4}}$},\hfill}
\newskip\Bigfill \Bigfill = 0pt plus 1000fill
\newcommand{\DpNameLast}[2]{\hbox{#1$^{\ref{#2}}$}\hspace{\Bigfill}}
%
%\small
\footnotesize
\noindent
\DpName{J.Abdallah}{LPNHE}
\DpName{P.Abreu}{LIP}
\DpName{W.Adam}{VIENNA}
\DpName{P.Adzic}{DEMOKRITOS}
\DpName{T.Albrecht}{KARLSRUHE}
\DpName{T.Alderweireld}{AIM}
\DpName{R.Alemany-Fernandez}{CERN}
\DpName{T.Allmendinger}{KARLSRUHE}
\DpName{P.P.Allport}{LIVERPOOL}
\DpName{U.Amaldi}{MILANO2}
% ...
\normalsize
\endgroup
\titlefoot{Department of Physics and Astronomy, Iowa State
     University, Ames IA 50011-3160, USA
    \label{AMES}}
\titlefoot{Physics Department, Universiteit Antwerpen,
     Universiteitsplein 1, B-2610 Antwerpen, Belgium \\
     \indent~~and IIHE, ULB-VUB,
     Pleinlaan 2, B-1050 Brussels, Belgium \\
     \indent~~and Facult\'e des Sciences,
     Univ. de l'Etat Mons, Av. Maistriau 19, B-7000 Mons, Belgium
    \label{AIM}}
\titlefoot{Physics Laboratory, University of Athens, Solonos Str.
     104, GR-10680 Athens, Greece
    \label{ATHENS}}
\titlefoot{Department of Physics, University of Bergen,
     All\'egaten 55, NO-5007 Bergen, Norway
    \label{BERGEN}}
\titlefoot{Dipartimento di Fisica, Universit\`a di Bologna and INFN,
     Via Irnerio 46, IT-40126 Bologna, Italy
    \label{BOLOGNA}}
\titlefoot{Centro Brasileiro de Pesquisas F\'{\i}sicas, rua Xavier Sigaud 150,
     BR-22290 Rio de Janeiro, Brazil \\
     \indent~~and Depto. de F\'{\i}sica, Pont. Univ. Cat\'olica,
     C.P. 38071 BR-22453 Rio de Janeiro, Brazil \\
     \indent~~and Inst. de F\'{\i}sica, Univ. Estadual do Rio de Janeiro,
     rua S\~{a}o Francisco Xavier 524, Rio de Janeiro, Brazil
    \label{BRASIL}}
\titlefoot{Coll\`ege de France, Lab. de Physique Corpusculaire, IN2P3-CNRS,
     FR-75231 Paris Cedex 05, France
    \label{CDF}}
\titlefoot{CERN, CH-1211 Geneva 23, Switzerland
    \label{CERN}}
\titlefoot{Institut de Recherches Subatomiques, IN2P3 - CNRS/ULP - BP20,
     FR-67037 Strasbourg Cedex, France
    \label{CRN}}
\titlefoot{Now at DESY-Zeuthen, Platanenallee 6, D-15735 Zeuthen, Germany
    \label{DESY}}
\titlefoot{Institute of Nuclear Physics, N.C.S.R. Demokritos,
     P.O. Box 60228, GR-15310 Athens, Greece
    \label{DEMOKRITOS}}
\titlefoot{FZU, Inst. of Phys. of the C.A.S. High Energy Physics Division,
     Na Slovance 2, CZ-180 40, Praha 8, Czech Republic
    \label{FZU}}
\titlefoot{Dipartimento di Fisica, Universit\`a di Genova and INFN,
     Via Dodecaneso 33, IT-16146 Genova, Italy
    \label{GENOVA}}
\titlefoot{Institut des Sciences Nucl\'eaires, IN2P3-CNRS, Universit\'e
     de Grenoble 1, FR-38026 Grenoble Cedex, France
    \label{GRENOBLE}}
\titlefoot{Helsinki Institute of Physics and Department of Physical Sciences,
     P.O. Box 64, FIN-00014 University of Helsinki, 
     \indent~~Finland
    \label{HELSINKI}}
\titlefoot{Joint Institute for Nuclear Research, Dubna, Head Post
     Office, P.O. Box 79, RU-101 000 Moscow, Russian Federation
    \label{JINR}}
\titlefoot{Institut f\"ur Experimentelle Kernphysik,
     Universit\"at Karlsruhe, Postfach 6980, DE-76128 Karlsruhe,
     Germany
    \label{KARLSRUHE}}
\titlefoot{Institute of Nuclear Physics PAN,Ul. Radzikowskiego 152,
     PL-31142 Krakow, Poland
    \label{KRAKOW1}}
\titlefoot{Faculty of Physics and Nuclear Techniques, University of Mining
     and Metallurgy, PL-30055 Krakow, Poland
    \label{KRAKOW2}}
\titlefoot{Universit\'e de Paris-Sud, Lab. de l'Acc\'el\'erateur
     Lin\'eaire, IN2P3-CNRS, B\^{a}t. 200, FR-91405 Orsay Cedex, France
    \label{LAL}}
\titlefoot{School of Physics and Chemistry, University of Lancaster,
     Lancaster LA1 4YB, UK
    \label{LANCASTER}}
\titlefoot{LIP, IST, FCUL - Av. Elias Garcia, 14-$1^{o}$,
     PT-1000 Lisboa Codex, Portugal
    \label{LIP}}
\titlefoot{Department of Physics, University of Liverpool, P.O.
     Box 147, Liverpool L69 3BX, UK
    \label{LIVERPOOL}}
\titlefoot{Dept. of Physics and Astronomy, Kelvin Building,
     University of Glasgow, Glasgow G12 8QQ
    \label{GLASGOW}}
\titlefoot{LPNHE, IN2P3-CNRS, Univ.~Paris VI et VII, Tour 33 (RdC),
     4 place Jussieu, FR-75252 Paris Cedex 05, France
    \label{LPNHE}}
\titlefoot{Department of Physics, University of Lund,
     S\"olvegatan 14, SE-223 63 Lund, Sweden
    \label{LUND}}
\titlefoot{Universit\'e Claude Bernard de Lyon, IPNL, IN2P3-CNRS,
     FR-69622 Villeurbanne Cedex, France
    \label{LYON}}
\titlefoot{Dipartimento di Fisica, Universit\`a di Milano and INFN-MILANO,
     Via Celoria 16, IT-20133 Milan, Italy
    \label{MILANO}}
\titlefoot{Dipartimento di Fisica, Univ. di Milano-Bicocca and
     INFN-MILANO, Piazza della Scienza 2, IT-20126 Milan, Italy
    \label{MILANO2}}
\titlefoot{IPNP of MFF, Charles Univ., Areal MFF,
     V Holesovickach 2, CZ-180 00, Praha 8, Czech Republic
    \label{NC}}
\titlefoot{NIKHEF, Postbus 41882, NL-1009 DB
     Amsterdam, The Netherlands
    \label{NIKHEF}}
\titlefoot{National Technical University, Physics Department,
     Zografou Campus, GR-15773 Athens, Greece
    \label{NTU-ATHENS}}
\titlefoot{Physics Department, University of Oslo, Blindern,
     NO-0316 Oslo, Norway
    \label{OSLO}}
\titlefoot{Dpto. Fisica, Univ. Oviedo, Avda. Calvo Sotelo
     s/n, ES-33007 Oviedo, Spain
    \label{OVIEDO}}
\titlefoot{Department of Physics, University of Oxford,
     Keble Road, Oxford OX1 3RH, UK
    \label{OXFORD}}
\titlefoot{Dipartimento di Fisica, Universit\`a di Padova and
     INFN, Via Marzolo 8, IT-35131 Padua, Italy
    \label{PADOVA}}
\titlefoot{Rutherford Appleton Laboratory, Chilton, Didcot
     OX11 OQX, UK
    \label{RAL}}
\titlefoot{Dipartimento di Fisica, Universit\`a di Roma II and
     INFN, Tor Vergata, IT-00173 Rome, Italy
    \label{ROMA2}}
\titlefoot{Dipartimento di Fisica, Universit\`a di Roma III and
     INFN, Via della Vasca Navale 84, IT-00146 Rome, Italy
    \label{ROMA3}}
\titlefoot{DAPNIA/Service de Physique des Particules,
     CEA-Saclay, FR-91191 Gif-sur-Yvette Cedex, France
    \label{SACLAY}}
\titlefoot{Instituto de Fisica de Cantabria (CSIC-UC), Avda.
     los Castros s/n, ES-39006 Santander, Spain
    \label{SANTANDER}}
\titlefoot{Inst. for High Energy Physics, Serpukov
     P.O. Box 35, Protvino, (Moscow Region), Russian Federation
    \label{SERPUKHOV}}
\titlefoot{J. Stefan Institute, Jamova 39, SI-1000 Ljubljana, Slovenia
     and Laboratory for Astroparticle Physics,\\
     \indent~~Nova Gorica Polytechnic, Kostanjeviska 16a, SI-5000 Nova Gorica, Slovenia, \\
     \indent~~and Department of Physics, University of Ljubljana,
     SI-1000 Ljubljana, Slovenia
    \label{SLOVENIJA}}
\titlefoot{Fysikum, Stockholm University,
     Box 6730, SE-113 85 Stockholm, Sweden
    \label{STOCKHOLM}}
\titlefoot{Dipartimento di Fisica Sperimentale, Universit\`a di
     Torino and INFN, Via P. Giuria 1, IT-10125 Turin, Italy
    \label{TORINO}}
%\titlefoot{INFN,Sezione di Torino, and Dipartimento di Fisica Teorica,
%     Universit\`a di Torino, Via P. Giuria 1,\\
%     \indent~~IT-10125 Turin, Italy
\titlefoot{INFN,Sezione di Torino and Dipartimento di Fisica Teorica,
     Universit\`a di Torino, Via Giuria 1,
     IT-10125 Turin, Italy
    \label{TORINOTH}}
\titlefoot{Dipartimento di Fisica, Universit\`a di Trieste and
     INFN, Via A. Valerio 2, IT-34127 Trieste, Italy \\
     \indent~~and Istituto di Fisica, Universit\`a di Udine,
     IT-33100 Udine, Italy
    \label{TU}}
\titlefoot{Univ. Federal do Rio de Janeiro, C.P. 68528
     Cidade Univ., Ilha do Fund\~ao
     BR-21945-970 Rio de Janeiro, Brazil
    \label{UFRJ}}
\titlefoot{Department of Radiation Sciences, University of
     Uppsala, P.O. Box 535, SE-751 21 Uppsala, Sweden
    \label{UPPSALA}}
\titlefoot{IFIC, Valencia-CSIC, and D.F.A.M.N., U. de Valencia,
     Avda. Dr. Moliner 50, ES-46100 Burjassot (Valencia), Spain
    \label{VALENCIA}}
\titlefoot{Institut f\"ur Hochenergiephysik, \"Osterr. Akad.
     d. Wissensch., Nikolsdorfergasse 18, AT-1050 Vienna, Austria
    \label{VIENNA}}
\titlefoot{Inst. Nuclear Studies and University of Warsaw, Ul.
     Hoza 69, PL-00681 Warsaw, Poland
    \label{WARSZAWA}}
\titlefoot{Fachbereich Physik, University of Wuppertal, Postfach
     100 127, DE-42097 Wuppertal, Germany
    \label{WUPPERTAL}}
\addtolength{\textheight}{-10mm}
\addtolength{\footskip}{5mm}
\clearpage
\headsep 30.0pt
\end{titlepage}
%%%%%%%%%%%%%%%%%%%%%%%%%
%
% Change for the document body
%%\pagestyle{heading} % for page numbering
\pagenumbering{arabic} % page numbering in number
\setcounter{footnote}{0} %
\large
%\linenumbers %%%CD
\section{Introduction}
\label{sec-intr}
%%%%%%%%%%%%%%%%%%%%%%%%%%%%%%%%%%%%%%%%%%%%%%%%%%%%%%%%%%%%%%%%%%%%%%%

In $e^+e^-$ collisions, data collected at high energies are
predominantly of hadronic nature showing a multi-jet final state
topology. At LEP, these data have led to measurements of many of the
Standard Model (SM) parameters and allowed limits to be set on new
physics processes. In some cases the original quark flavour and its
mass have not been a critical issue in performing the measurement and
therefore approximations using massless quarks have been sufficient
for the required precision. This is well justified for inclusive-type
observables like total cross-sections for which the correction due to
massive quarks depends on the quark mass, $m_q$, and on the energy of
the process, $Q$, as $m_q^2c^4/Q^2$. For $b$ quarks ($m_q=m_b\sim 3-5$
GeV/$c^2$) at LEP I centre-of-mass energies ($Q=M_Zc^2$) this
represents an effect of less than three per mille. On the other hand,
for more exclusive observables, such as multi-jet cross-sections, the
mass dependence transforms into terms proportional to $m_q^2c^4/(Q^2
\cdot y_c)$ where $y_c$, the jet resolution variable, usually takes
values much lower than unity, enabling these mass effects to become
sizeable \cite{proposta_mbmz}. Differences in the multi-jet production
rate for massive $b$ quarks with respect to the corresponding rate for
massless $\ell$ quarks ($\ell \equiv u,d,s$) can then be as large as
3\% to 20\% for three- or four-jet final states.

The large volume of data collected by the LEP experiments and the
highly-effective techniques developed to identify the quark flavour of
the jets, have increased the experimental sensitivity for observables
where the quark flavour is relevant \cite{delphi_btag}. Consequently
theoretical predictions including mass corrections have become
necessary to reach a proper understanding of such observables in order
to interpret them as standard or new physics. In some cases these mass
effects have only been computed at leading order (LO) but for some
event-shape variables and in particular for the three-jet rate,
calculations including next-to-leading order (NLO) terms also exist
\cite{nlo1,nlo2,nlo3,nlo4}. Rather precise experimental studies on the
production of multi-jet events initiated by $b$ quarks have then been
allowed and performed at LEP and SLC
\cite{delmbmz,sldmbmz,alephmb,opalasb,opalmb}. Results obtained agree
well with the predictions of the Standard Model, i.e., Quantum
Chromodynamics (QCD). The quantification of these mass effects has
allowed a verification of the flavour independence of the strong
coupling constant to a precision of less than 1\% and an extraction of
a value of the $b$ mass at the energy scale of the $Z$ boson mass,
$Q=M_Zc^2$, within an uncertainty of $\sim0.4-0.5$ GeV/$c^2$. These
measurements have in addition provided the first evidence of the
running of the $b$ mass, i.e. the evolution of this parameter as a
function of the energy scale when compared to the values obtained from
processes occuring at lower energy scales.

%However this latter statement  
%still remains under debate \cite{nason_roma}.

A reduction of the uncertainty of the $b$ mass determination could be
accomplished by a combination of the present individual LEP and SLC
results. This is certainly difficult and, at the end may improve the
experimental precision only slightly, because the dominant errors are
of systematic nature and common to all measurements. Hence, the best
way to significantly increase the accuracy for this parameter is by a
deeper understanding of the physics processes and correction
procedures involved in the analyses.

%%%%%%%%%%%%%%%%%%%%%%%%%%%%%%%%%%%%%%%%%%%%%%%%%%%%%%%%%%%%%%%%%%%%%%%
\subsection{The $b$ quark mass and the observable}
\label{subsec-intr}
%%%%%%%%%%%%%%%%%%%%%%%%%%%%%%%%%%%%%%%%%%%%%%%%%%%%%%%%%%%%%%%%%%%%%%%

The $b$ quark mass is a free parameter in the SM Lagrangian and
therefore needs to be measured experimentally. Precise determinations
of this parameter are very interesting both as a fundamental parameter
and also to constrain models beyond the SM. Unfortunately, confinement
of quarks inside the observed hadrons introduces additional
complications not present in mass determinations of free particles as
for instance leptons. Quark masses need then to be defined within a
theoretical convention and can only be inferred indirectly through
their influence on hadronic observables.

Among the different quark mass definitions, the most commonly used in
high energy processes are the pole mass $M_b$ and the running mass
$m_b(Q)$. The former is defined as the pole of the renormalized quark
propagator and is gauge and scheme independent. The latter corresponds
to the renormalized mass in the $\overline{\rm MS}$ scheme and depends
on the process energy ($Q$). These mass definitions are related to
each other \cite{tarrach} and NLO calculations are needed in order to
distinguish between the two. Physics is independent of the mass
definition. However, when using perturbation theory at a fixed order
to extract physics results a dependence appears on the mass definition
as well as on the arbitrary renormalization scale $\mu$.  This is due
to neglected perturbative and non-perturbative higher order terms and
therefore it is possible than one scheme might be more convenient than
another for a given purpose \cite{bmass_02,hoang_02}.  Specially
because the relation between the two mass definitions does not have a
good convergence behaviour due to renormalon ambiguities
\cite{hep-ph/9505317}.

In the case of the running mass $m_b(Q)$, the largest part of its
running occurs at low energy scales up to $M_Z/2$. The exact mass
running represents a basic constraint to theories beyond the SM, such
as those implying the unification of the $b$ quark and $\tau$ lepton
masses at the Grand Unified Theory (GUT) scale. The $b$ quark mass has
also important implications on Higgs searches since the partial decay
width of the Higgs boson into $b$ quarks is proportional to the $b$
quark mass squared. In this case, it can be shown that the mass
definition is also relevant for the accuracy of the theoretical
prediction \cite{higgs1,higgs2,higgs3}.

At low energy, the $b$ quark mass is established from the measured
spectra of hadronic bound states or the moments of the spectrum of the
$B$ decay products making use of non-perturbative techniques such as
the Heavy Quark Effective Theory (HQET), Non-Relativistic QCD (NRQCD),
QCD sum rules or lattice QCD. An attempt to average these
determinations properly is presented in \cite{bmass_02} obtaining the
value $m_b(m_b) = 4.24 \pm 0.11$ GeV/$c^2$.

At high energy the $b$ quark mass has been extracted from data
collected at the $Z$ peak at LEP and SLC. The first measurement of
$m_b(M_Z)$ was performed by the {\sc Delphi} experiment with data
collected from 1992 to 1994 at $\sqrt{s} \approx M_Z$. The observable
used in this analysis was:
\begin{equation}
R_3^{b\ell}(y_c) =
\frac{R_3^b (y_c)}{R_3^\ell (y_c)} =
\frac{\Gamma_{3j}^b(y_c)/\Gamma^b}{\Gamma_{3j}^\ell(y_c)/\Gamma^\ell}
\label{eq:obs}
\end{equation}
with $\Gamma_{3j}^q(y_c)$ and $\Gamma^q$ being, respectively, the
three-jet and total decay widths of the $Z$ into $q\overline{q}$,
where $q$ = $b$ or $\ell$ ($\ell \equiv u,d,s$ quarks). The flavour
$q$ of the hadronic event was defined as that of the quarks coupling
to the $Z$ and the {\sc Durham} algorithm was used for the jet
clustering. The measured observable was compared with the theoretical
computations of \cite{nlo1} and $m_b$ in the $\overline{\rm MS}$
scheme was found to be \cite{delmbmz}:
\begin{equation}
m_b(M_Z) = 
2.67 \pm 0.25~({\rm stat}) \pm 0.34 ~({\rm had}) \pm 0.27~({\rm theo})~{\rm
GeV/}c^2 
\label{eq:mb_delphi_durham}
\end{equation}
where the quoted error was mainly due to the hadronization uncertainty.

In this paper a new analysis to measure the $b$ mass, performed with
data collected by {\sc Delphi} from 1994 to 2000 is presented. The
data taken in the years before have not been considered because the
Vertex Detector layout was changed in 1994, improving its capability
since then (see Section \ref{sec-delphi}). The same observable as used
in the previous {\sc Delphi} measurement, $R_3^{b\ell}$, has been used
with two jet-clustering algorithms, {\sc Cambridge} \cite{cambridge}
and {\sc Durham} \cite{durham}. The {\sc Cambridge} algorithm has the
advantage of having a smaller theoretical uncertainty \cite{nlo4}. A
detailed study of how mass effects and the hadronization process are
implemented in the fragmentation models has led to a better control of
the hadronization correction. The effect of the gluon-splitting rates
into $b$ and $c$ quarks on the flavour tagging has also been taken
into account.

%%%%%%%%%%%%%%%%%%%%%%%%%%%%%%%%%%%%%%%%%%%%%%%%%%%%%%%%%%%%%%%%%%%%%%%%%%
\section{The {\sc Delphi} detector}
\label{sec-delphi}
%%%%%%%%%%%%%%%%%%%%%%%%%%%%%%%%%%%%%%%%%%%%%%%%%%%%%%%%%%%%%%%%%%%%%%%%%%

A brief description of the most relevant components of the {\sc
Delphi} detector for this analysis is given here. A detailed
description of its design and performance can be found in
\cite{delphi1,delphi2}.

{\sc Delphi} was a hermetic detector with a superconducting solenoid
providing a uniform magnetic field of 1.23 T parallel to the beam axis
throughout the central tracking device volume.

The tracking system consisted of a silicon Vertex Detector (VD), a jet
chamber Inner Detector (ID) and a Time Projection Chamber (TPC) which
constituted the main tracking device in {\sc Delphi}. At a larger
distance from the interaction point the tracking was complemented by a
drift chamber Outer Detector (OD) covering the barrel region
($40^\circ \le \theta \le 140^\circ$) and two sets of drift chambers,
FCA and FCB, located in the endcaps.

The VD was made of three coaxial cylindrical layers of silicon
strips. From 1994 onwards the outer and innermost layers were equiped
with doubled-sided detectors with orthogonal strips, allowing the
measurement of both $R\phi$ and $z$ coordinates. In 1996, the VD was
doubled in length and in 1997 a Very Forward Detector consisting of
ministrips and pixels was added. Earlier {\sc Delphi} data taken in
periods with a less complete VD setup are not included in this
analysis whereas data collected, later, in the period of LEP2,
1996-2000, which corresponded to the calibration runs at the
centre-of-mass energy of the $Z$ peak are used.

Electron and photon identification was provided mainly by the
electromagnetic calorimeter which was composed of a High Density Projection
Chamber (HPC) installed inside the coil in the barrel region and a lead-glass
calorimeter (FEMC) in the forward region.  

In order to measure the charged and neutral hadronic energy, {\sc Delphi}
also included the hadron calorimeter (HCAL), an iron-gas sampling detector
incorporated in the magnet yoke. 

%%%%%%%%%%%%%%%%%%%%%%%%%%%%%%%%%%%%%%%%%%%%%%%%%%%%%%%%%%%%%%%%%%%%%%%%%%%
\section{Hadronization correction of $R_3^{b\ell}$}
\label{sec-chad}
%%%%%%%%%%%%%%%%%%%%%%%%%%%%%%%%%%%%%%%%%%%%%%%%%%%%%%%%%%%%%%%%%%%%%%%%%%%

The hadronization correction to $R_3^{b\ell}$, i.e. the ratio of the
observable at parton over hadron level, was computed in \cite{delmbmz}
using the string-fragmentation and cluster models implemented in {\sc
Jetset} 7.3 \cite{jetset} and {\sc Herwig 5.8} \cite{herwig58}
respectively, previously tuned to {\sc Delphi} data
\cite{delphi_tuning}.  Uncertainties coming from the tuned parameters
of {\sc Jetset} and from the different predictions of the two
fragmentation models were taken into account, the latter being the
highest contribution to the total error.

In the present analysis more recent versions of the generators in
which $b$ quark mass effects are better modelled ({\sc Pythia} 6.131
\cite{pythia} and {\sc Herwig} 6.1 \cite{herwignou}) were used.  For
the case of {\sc Pythia}, different fragmentation functions were
considered: Peterson \cite{peterson} and Bowler \cite{bowler}
\footnote{The {\sc Pythia} parameters used are PARJ(55)=-0.00284 for
Peterson and MSTJ(11)=5, PARJ(46)=1 and PARJ(47)=0.95 for Bowler. }.  The
final hadronization correction applied to our observable was the one
given by {\sc Pythia} 6.131 with the Peterson $b$ fragmentation
function since this model gives the best overall description of the
{\sc Delphi} data and the other two cases were only used to evaluate
the model uncertainty.

The model uncertainty was reduced to a negligible effect by performing 
the measurement in a restricted region of the phase space. 
Other sources of uncertainties such as the effect of the $b$ quark mass 
parameter used in the generator were studied in detail and were shown also to
be important. These two questions are discussed in the following sections.

%%%%%%%%%%%%%%%%%%%%%%%%%%%%%%%%%%%%%%%%%%%%%%%%%%%%%%%%%%%%%%%%%%%%%%%%%%%%
\subsection{Hadronization model uncertainty}
\label{sec-model}
%%%%%%%%%%%%%%%%%%%%%%%%%%%%%%%%%%%%%%%%%%%%%%%%%%%%%%%%%%%%%%%%%%%%%%%%%%%%

The hadronization model uncertainty, $\sigma_{mod}$, was evaluated as the 
standard deviation of the hadronization corrections predicted by
the cluster model implemented in {\sc Herwig} and the string-fragmentation
model of {\sc Pythia} using two heavy quark fragmentation functions, Peterson 
and Bowler. 

\begin{figure}
\begin{center}
  \includegraphics[width=0.50\linewidth]{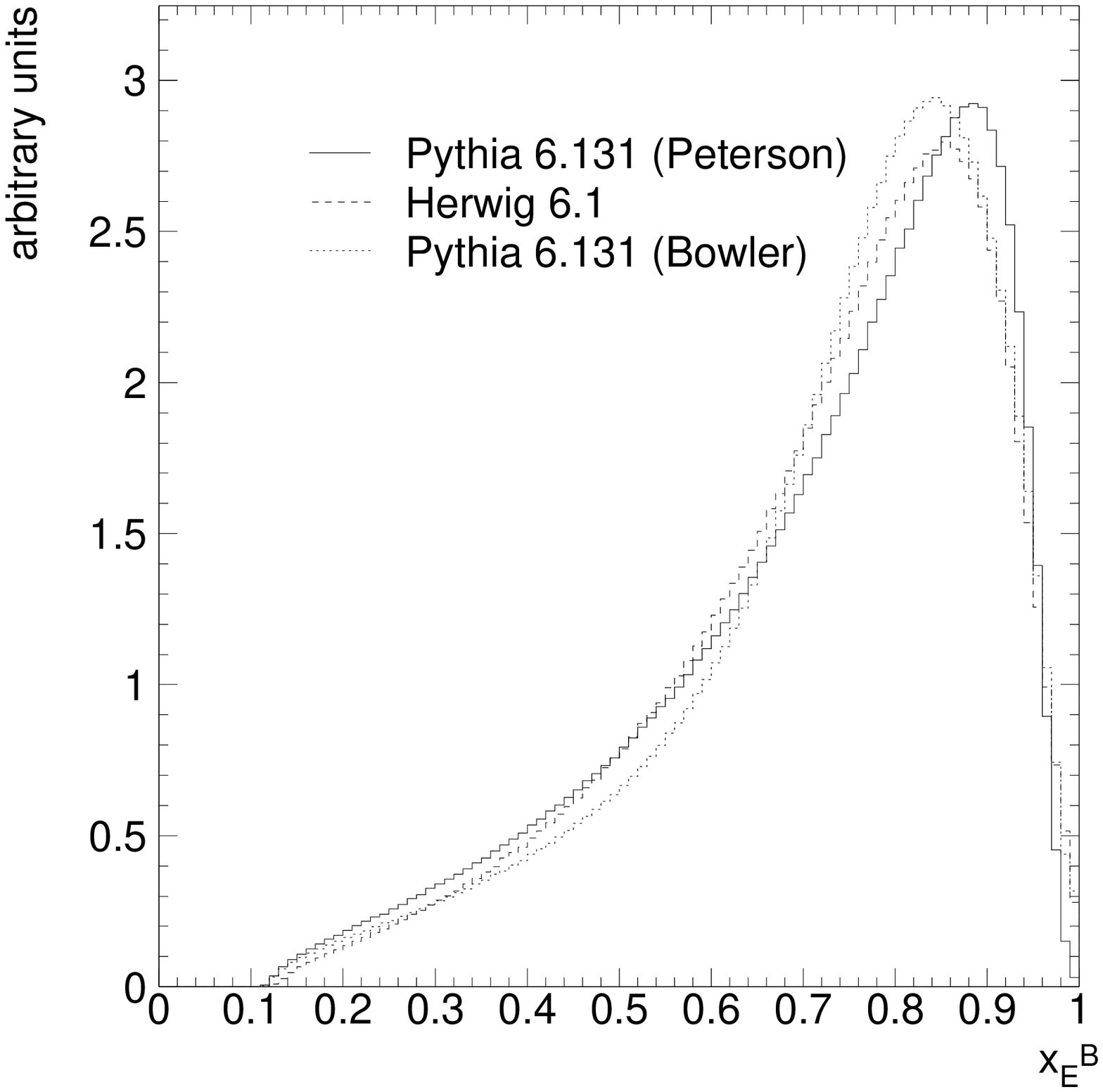} 
  \includegraphics[width=0.47\linewidth]{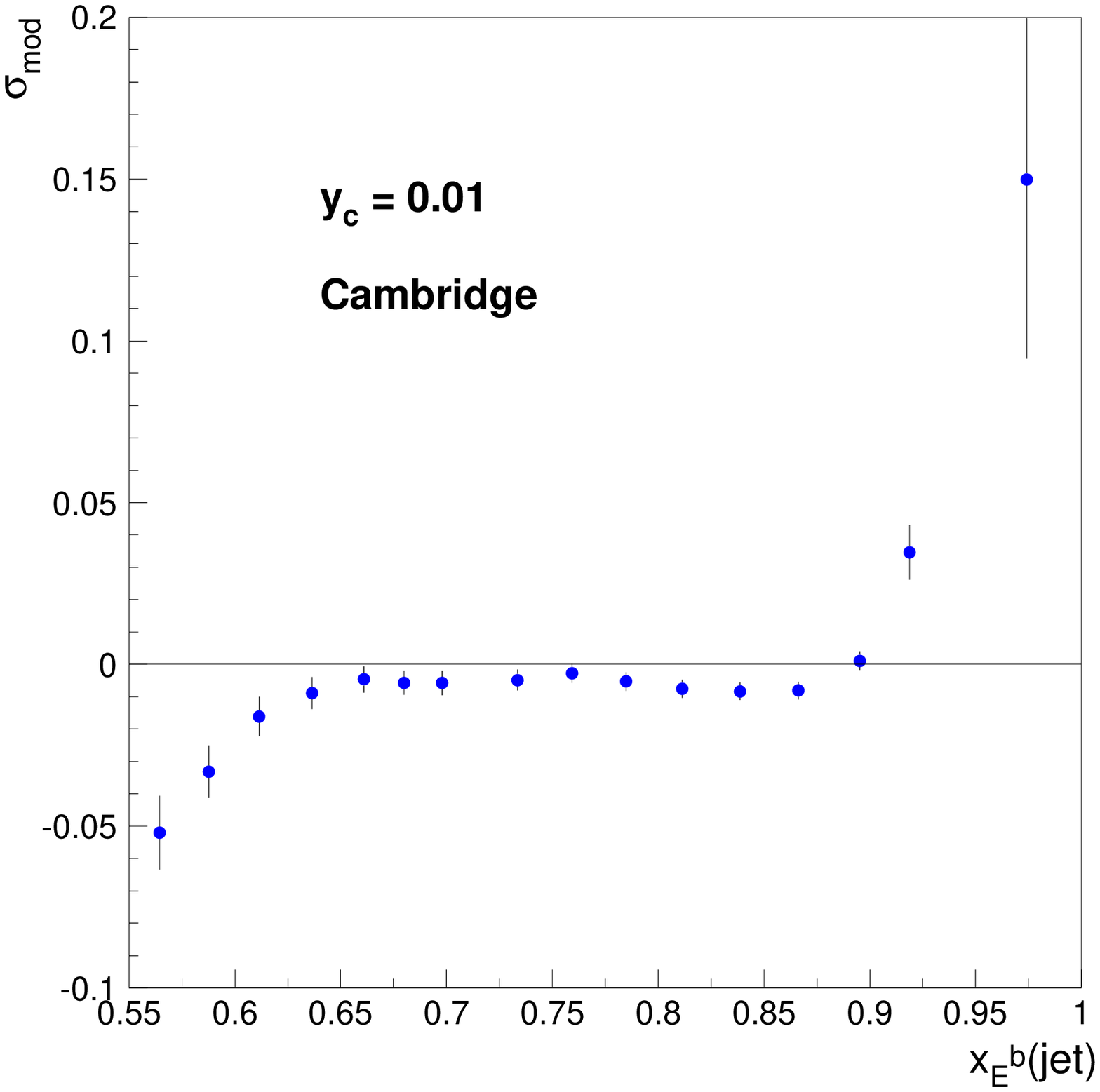} 
\end{center} 
\caption{${\rm x_E^B}$ distribution for the different Monte Carlo
generators (left). Hadronization model uncertainty (i.e. standard deviation 
of the hadronization corrections predicted by {\sc Herwig} and {\sc Pythia} 
with Peterson and Bowler heavy fragmentation functions)  
as a function of the mean of the ${\rm x_E^b}({\rm jet})$ that is defined 
in the text and for $y_c=0.01$ for {\sc Cambridge} (right).}
\label{fig:frag}
\end{figure}

\begin{figure}
\begin{center}
  \includegraphics[width=0.45\linewidth]{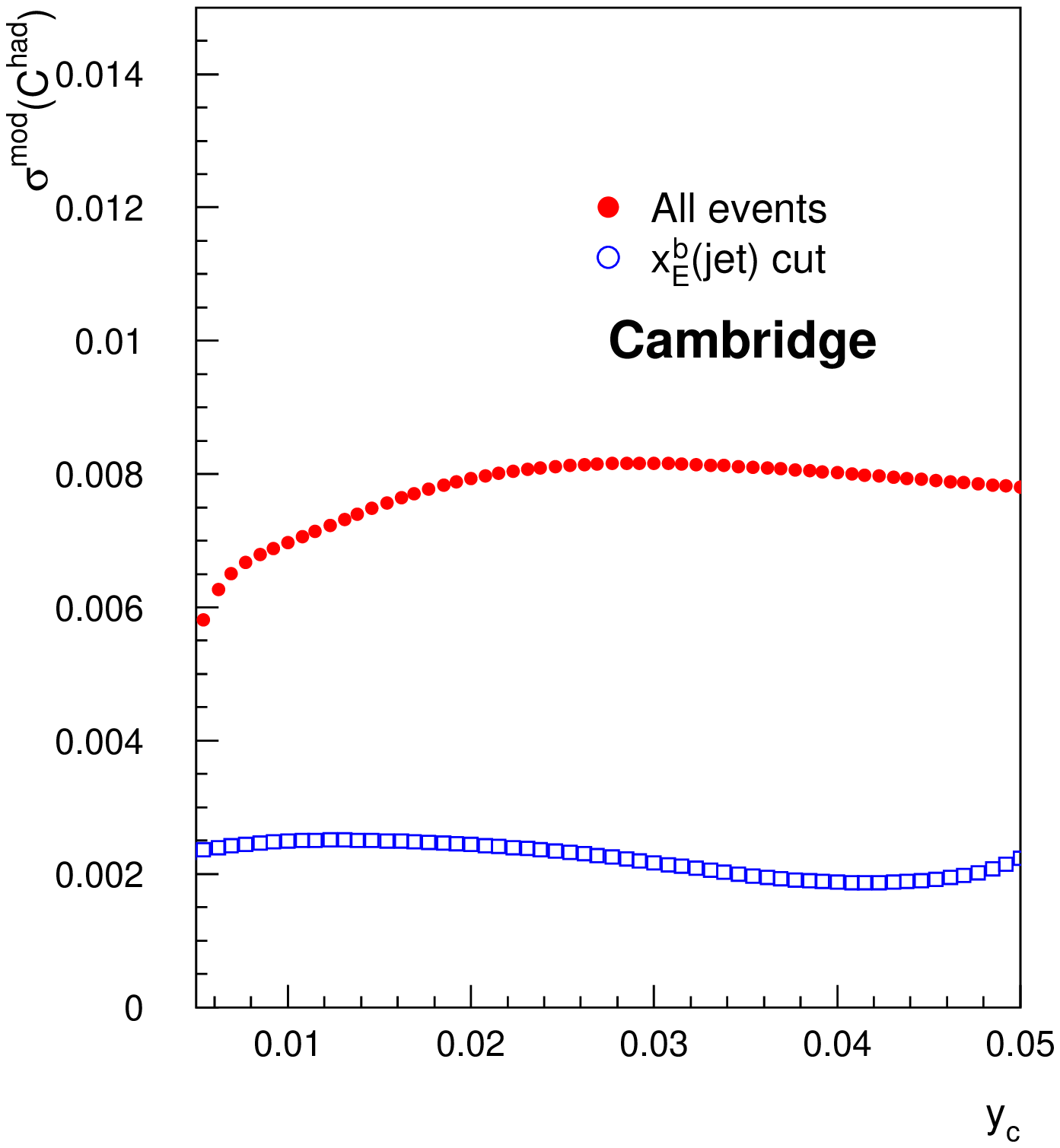} 
  \includegraphics[width=0.45\linewidth]{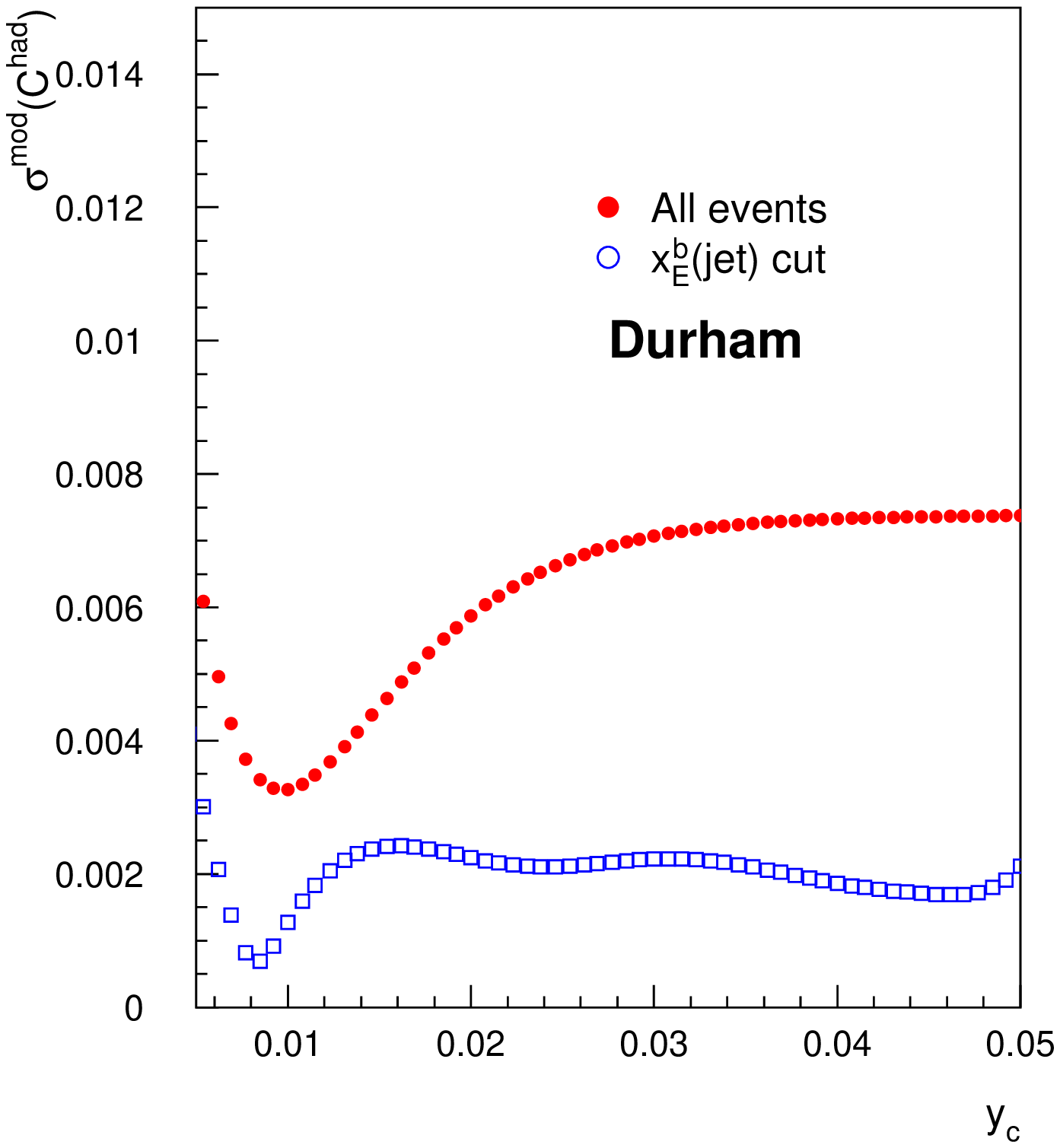} 
\end{center} 
\caption{Hadronization model uncertainty as a function of the $y_c$ for the
{\sc Cambridge} (left) and {\sc Durham} (right) algorithms.}
\label{fig:emodel}
\end{figure}

The hadronization corrections were found to depend on the $B$-hadron scaled 
energy, ${\rm x_E^B} = 2 E_{B-{\rm hadron}}/M_Z$, the distribution of which is 
shown in Figure \ref{fig:frag}. As this quantity and the corresponding jet 
energy including the $B$-hadron are highly correlated a new variable was 
defined instead: the $b$-jet scaled energy 
${\rm x_E^b}({\rm jet}) = 2 E_{b-\mathrm{jet}}/M_Z$ where 
$E_{b-\rm{jet}}$ is the energy of the jets originated by the primary
$b$ quarks in a $Z \rightarrow b\overline{b}$ event. The study of the 
dependence of the hadronization corrections with ${\rm x_E^b}({\rm jet})$ led
to the conclusion that if the cut
${\rm x_E^b}(\mathrm{jet}) \ge 0.55$ is applied to both $b$ quark jets 
the model uncertainty is reduced
by a factor of 4 (see the right hand plot in Figure \ref{fig:frag} and 
Figure \ref{fig:emodel} ).

%%%%%%%%%%%%%%%%%%%%%%%%%%%%%%%%%%%%%%%%%%%%%%%%%%%%%%%%%%%%%%%%%%%%%%%%%%%%
\subsection{The $b$ mass parameter in the generator}
\label{sec-masspythia}
%%%%%%%%%%%%%%%%%%%%%%%%%%%%%%%%%%%%%%%%%%%%%%%%%%%%%%%%%%%%%%%%%%%%%%%%%%%%

In this section, the effect of the $b$ quark mass parameter used in 
{\sc Pythia} on the hadronization correction is 
discussed. The result of this study also applies to other generators 
which contain similar features.

In order to describe $b$ quark mass effects, {\sc Pythia} uses a set of
three $b$ quark mass parameters: the kinematical mass, $M_b^{kine}$, used in the parton 
shower (PS) process,
the constituent mass, $M_b^{const}$, used during the hadronization
process and finally the known $B$ hadron masses. The constituent mass is also
used to derive masses for predicted but not yet observed $B$ hadrons. 
In the model these 
three masses are not connected to each other and, as a consequence, mass 
effects at parton level do not automatically propagate to the 
hadronization process, as they physically should. This results in a
dependence of the hadronization correction on the $b$ quark mass of the 
generator. 

If the various mass parameters are connected by, for instance, making the 
constituent and kinematical $b$ quark masses equal to each other and by 
deriving all $B$ hadron masses from the corresponding quark masses 
using the hadron mass formula \cite{derujula}, this dependence of the 
hadronization correction factor is completely removed. This feature 
of the generators was also noticed in previous studies 
\cite{delmbmz,sldmbmz,alephmb,opalasb,opalmb} even though the exact cause 
of this behaviour was not identified. Unfortunately, this argument
cannot define the value to be assigned to the mass parameters of the standard generator
in which the quark masses are not connected. For that purpose, the following 
procedure was applied:

\begin{itemize}
\item[$\bullet$]
In order to assess the precision of the massive calculation implicit in the 
parton shower generator, its prediction for $R_3^{b\ell}$ and that of the 
NLO calculation was first compared. The method was to change the input mass in
the NLO calculations to minimize their overall difference. Then the difference 
of the input mass values was evaluated. For the parton shower the so called 
kinematical mass, $M_b^{kine}$, was employed and for the NLO calculations the two 
mass definitions were considered. In the case of the running mass the corresponding
value was transformed to the pole mass 
\footnote{The 3-loop relationship between $m_b(m_b)$ and $M_b$ \cite{mb-Mb}
with $\alpha_s(M_Z) = 0.1183 \pm 0.0027$ \cite{alphas} was used.}. 
The value $\Delta (M_b^{kine}-M_b) \sim 15$ MeV/$c^2$ was obtained in the 
case of the running mass and $\Delta (M_b^{kine}-M_b) \sim 500$ MeV/$c^2$ for the
pole mass. These differences were later considered as the uncertainty 
associated to the effective mass definition of the parton shower.

\item[$\bullet$]
The values of the $b$ quark mass measurements determined from low energy processes were then used 
as input to the mass parameter, $M_b^{kine}$, of the generator entering in 
the hadronization process. In order to 
select the mass value to be used in the present analysis, various possibilites were explored. A direct 
determination of $M_b$ from 
reference \cite{markus} gave $M_b = 4.98 \pm 0.13$ GeV/$c^2$. A second 
possibility is to use the average value for the running 
mass calculated in \cite{bmass_02} as: $m_b(m_b) = 4.24 \pm 0.11$ GeV/$c^2$, 
which could be 
transformed into a pole mass value of $M_b = 4.99 \pm 0.13$ GeV/$c^2$. 
A third value is also available using {\sc Delphi} data from the semileptonic 
$B$ decays for which the relevant scale is that of the $B$ hadron masses, 
leading to $M_b = 5.00 \pm 0.16$ GeV/$c^2$ \cite{lambda_delphi}. All these 
results are compatible and have similar accuracy. The value which was used in
the generator to compute the hadronization correction was that obtained as
the average of all low energy measurements \cite{bmass_02}:

\begin{equation}
m_b(m_b) =  4.24 \pm 0.11 \ {\rm GeV}/c^2, \ \ {\rm or}, \ \ M_b = 4.99 \pm 0.13 \ {\rm GeV}/c^2.
\label{eq:mb_delphi_durham2}
\end{equation}
\end{itemize}

For cross-check purposes, values of the $b$ quark mass were also extracted using {\sc Delphi} data 
alone with the modified generator for which 
the set of the three $b$ quark masses are connected to each other. Two different observables were 
employed for this study: the $y_{32}$ distribution\footnote{$y_{32}$ is the $y_c$ transition 
value in which a 3-jet event becomes a 2-jet} of $b$ over $\ell$ events normalized to the total 
number of $b$ and $\ell$ events, $R^{b\ell}(y_{32})$ and the minimum angle between $b$ quark and gluon jets 
when every event is forced to three jets. Both quantities are correlated with the 
observable $R_3^{b\ell}$ used to measure the $b$ quark mass and therefore their role in the present analysis
is limited to qualitative checks. The first observable gave a fitted value for the $b$ quark mass
of the modified generator of $M_b = 4.93 \pm 0.13$ GeV/$c^2$ and the second one 
$M_b = 4.95 \pm 0.11$ GeV/$c^2$. These results are thus consistent with the choice of the mass parameter and 
the above quoted uncertainty.

%Finally, the kinematical $b$ quark mass parameter used in the parton shower evolution was set to 
%the same value of the $b$ pole mass: $M_b = 4.99 \pm 0.13 \ {\rm GeV}/c^2$ as for the
%hadronization process. The uncertainty due to the mass 
%identification, $M_k(b) \equiv M_b \pm \Delta (M_k(b)-M_b)$, was taken into account later 
%as an additional source of systematic errors in the determination of the $b$ mass. 

%%%%%%%%%%%%%%%%%%%%%%%%%%%%%%%%%%%%%%%%%%%%%%%%%%%%%%%%%%%%%%%%%%%%%%%%%%%
\section{Experimental determination of $R_3^{b\ell}$}
\label{sec-exp}
%%%%%%%%%%%%%%%%%%%%%%%%%%%%%%%%%%%%%%%%%%%%%%%%%%%%%%%%%%%%%%%%%%%%%%%%%%%

First the sample of $Z$ hadronic decays, i.e. $Z
\rightarrow q\overline{q}$ events was selected. Then the $b$ and $\ell$
quark-initiated events were separated using the {\sc Delphi} flavour tagging
methods and later a cut on the $b$ quark jet energy was also performed in order to 
discard those events with large hadronization correction (see Section 
\ref{sec-model}).

The jet-clustering algorithms {\sc Cambridge} and {\sc Durham}
were applied to both tagged samples to obtain the $R_3^{b\ell}$ observable at
detector level. Data were then corrected for detector and tagging effects and
for the hadronization process to bring the observable to parton level.

%%%%%%%%%%%%%%%%%%%%%%%%%%%%%%%%%%%%%%%%%%%%%%%%%%%%%%%%%%%%%%%%%%%%%%%%%%%
\subsection{Event selection}
\label{sec-sel}
%%%%%%%%%%%%%%%%%%%%%%%%%%%%%%%%%%%%%%%%%%%%%%%%%%%%%%%%%%%%%%%%%%%%%%%%%%%

The selection of $Z$ hadronic events was done in three steps (as in
\cite{delmbmz}): 

\begin{itemize}

  \item particle selection: Charged and neutral particles were selected
  in order to ensure a reliable determination of their momenta and energies
  by applying the cuts listed in Table \ref{tab:had_cuts}; 

  \item event selection: $Z \rightarrow q\overline{q}$ events were selected
  by imposing the global event conditions of Table \ref{tab:had_cuts};

  \item kinematic selection: In order to reduce particle losses and imperfect
  energy-momentum assignment to jets in the selected hadronic events, further
  kinematical cuts were applied. Each event was clustered into three jets by
  the jet-clustering algorithm ({\sc Cambridge} and {\sc Durham})
  using all selected charged and neutral particles. The cuts of Table
  \ref{tab:had_cuts} were then applied.

\end{itemize}

After applying these cuts to the data a sample of $1.4 \times 10^6$
($1.3 \times 10^6$) hadronic $Z$ decays was selected for the {\sc Cambridge}
({\sc Durham}) algorithm.

\begin{table}[thbp]
\centering
\vspace{7mm}
\begin{tabular}{|l|l|}
\hline
               & $p\geq$ 0.1 GeV/$c$ \\
Charged        & 25$^\circ \leq  \theta \leq 155 ^\circ$  \\
Particle       & L$\geq$ 50 cm                            \\
Selection~~~~~ & $d \leq$  5 cm in $R\phi$ plane    \\
               & $d \leq$ 10 cm in $z$ direction    \\
\hline
\hline

Neutral        & $E \geq 0.5$ GeV, ~40$^\circ$
                                     $ \leq  \theta  \leq $
                                     $140^\circ$                        
                                     (HPC) \\
Cluster        & $E \geq 0.5$ GeV, ~8$^\circ$(144$^\circ$)
                                     $ \leq  \theta  \leq $
                                     $36^\circ$(172$^\circ$) ~~~~~    
                                     (FEMC) \\
Selection~~~~~ & $E \geq 1$ GeV, ~10$^\circ \leq  \theta  \leq 170^\circ$
                                     (HCAL) \\
\hline
\hline
               & $N_{ch} \geq 5$                                     \\
               & $E_{ch} \geq 15$ GeV                                \\
Event          & $|\sum_i{q_i}| \leq 6$, $i=1,...,N_{ch}$            \\
Selection~~~~~ & No charged particle with $p \geq 40$ GeV/$c$        \\
               & 45$^\circ \leq  \theta_{thrust} \leq 135 ^\circ$    \\
\hline
\hline
               & $N^{ch}_j \geq 1$ per jet                                  \\
Kinematic      & $E_{j} \geq 1$ GeV,~ $j=1,2,3$                           \\
Selection~~~~~ & 25$^\circ \leq  \theta_{j} \leq 155 ^\circ$,~ $j=1,2,3$  \\
& Planarity cut: $\sum_{ij}{\phi_{ij}} \geq 359^\circ, \ \ \
i<j, \ \ i,j=1,2,3$\\\hline 
\end{tabular}
\vspace{0.5cm}
\caption{Particle and hadronic-event selection cuts; $p$ is the particle momentum,
$\theta$ the particle polar angle and $\theta_{thrust}$ the $thrust$ polar angle
(with respect to the beam axis in both cases), L the  measured track length, $d$ the closest
distance to the interaction point, $q_i$  the particle charge, $E$ the
cluster energy, $N_{ch}$ the number of charged  particles, and $E_{ch}$ the
total charged-particle energy in the event. The kinematic selection is based 
on the properties of the events when
clustered into three jets by the jet algorithm. $N^{ch}_j$ is the jet charged
multiplicity, $E_j$ the jet energy, $\theta_j$ the jet polar angle and
$\phi_{ij}$ the angular separation between the pair of jets $ij$.}
\label{tab:had_cuts}
\end{table}

%%%%%%%%%%%%%%%%%%%%%%%%%%%%%%%%%%%%%%%%%%%%%%%%%%%%%%%%%%%%%%%%%%%%%%%%%%%%
\subsection{Flavour tagging}
\label{sec-tag}
%%%%%%%%%%%%%%%%%%%%%%%%%%%%%%%%%%%%%%%%%%%%%%%%%%%%%%%%%%%%%%%%%%%%%%%%%%%%

The $b$ and light ($\ell \equiv u,d,s$) quark-initiated events need to
be identified. {\sc Delphi} has developed two different algorithms for
$b$ tagging based on those properties of $B$ hadrons that differ from
those of other particles: the impact parameter \cite{ip} and the
combined technique \cite{comb}. The former makes use of the most
important property for the selection of $B$ hadrons, their long
lifetime, and discriminates the flavour of the event by calculating
the probability, $P_E^+$, of having all particles compatible with
being generated at the event interaction point.  The second technique,
besides the impact parameter of charged particles, uses other
discriminating variables: the transverse momentum of any identified
energetic lepton with respect to the jet direction and, in case a
secondary vertex is found, the total invariant mass, the fraction of
energy, the transverse momentum and the rapidities of the charged
tracks belonging to the secondary vertex. An optimal combination of
this set of variables defined for each reconstructed jet is performed,
leading to a single variable per event, $X_{effev}$. When the aim was
to measured $R_3^{b\ell}$ for {\sc Cambridge}, this was also the
algorithm used to compute the combined tagging variable, and the same
for the case of {\sc Durham}.

\begin{figure}
\begin{center}
    \includegraphics[width=0.47\linewidth]{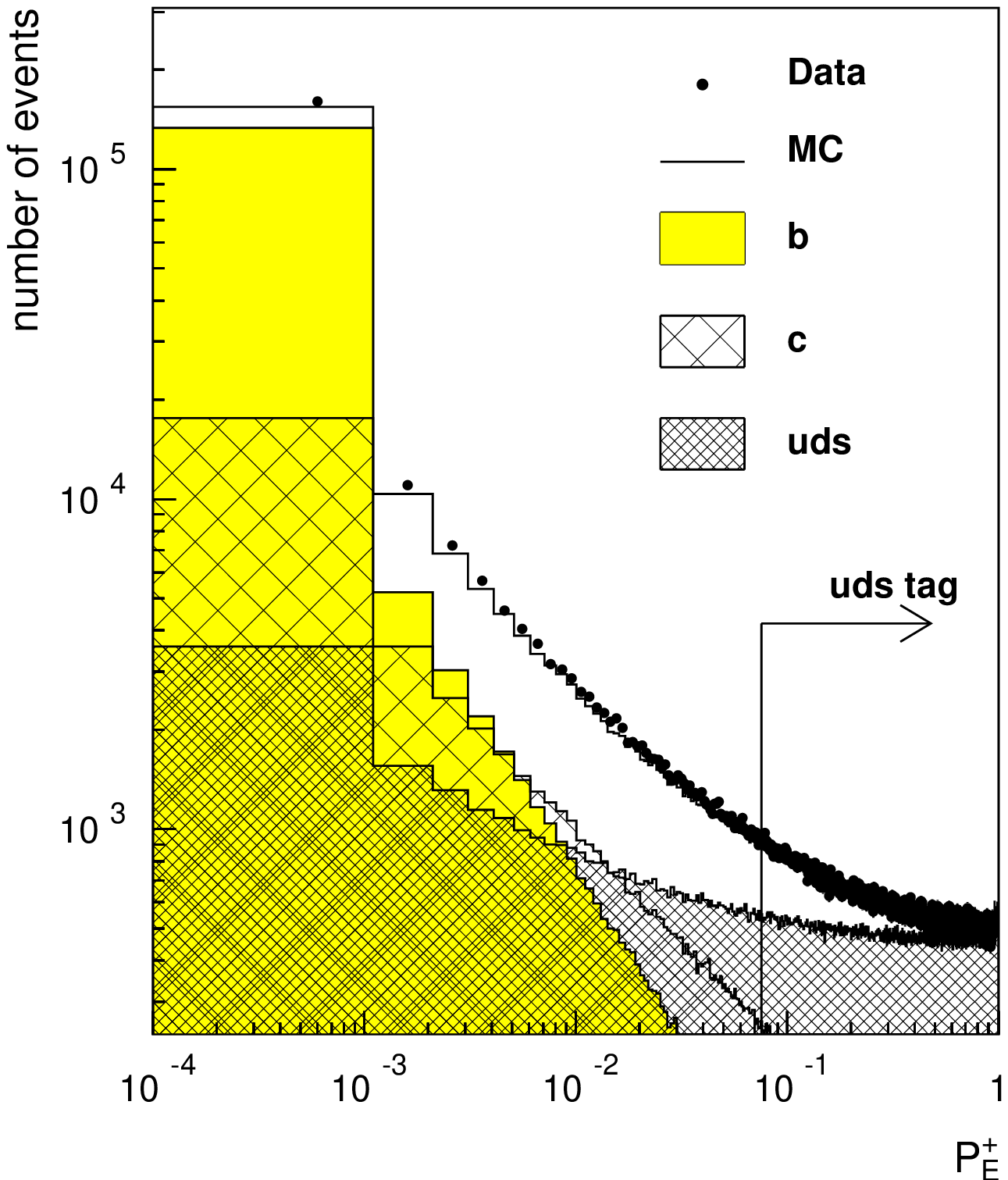} 
    \includegraphics[width=0.50\linewidth]{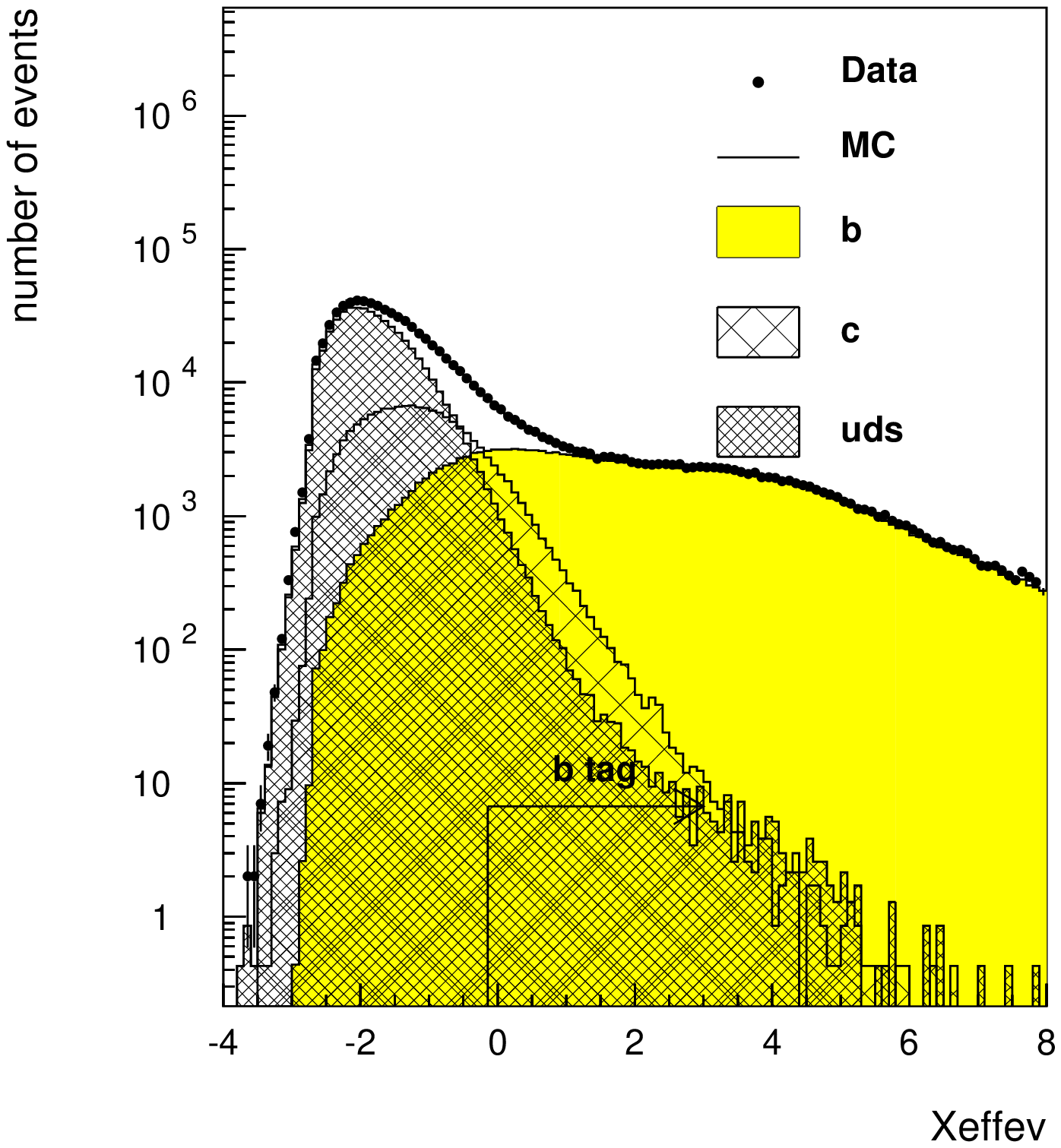} 
\end{center} 
\caption{Event distribution of $P_E^+$ (left) and $X_{effev}$ (right) when {\sc
Durham} is used to form jets.  The real (points) and simulated (histogram)
data are compared. The specific contribution of each quark flavour is
displayed as derived from the {\sc Delphi} simulated data. The cuts used to
tag the $b$ and $\ell$ quark sample are also indicated.}
\label{fig:ip_comb_tag}
\end{figure}

Figure \ref{fig:ip_comb_tag} shows the distributions of $P_E^+$ and
$X_{effev}$ obtained for the selected real and simulated sample of $Z$
hadronic decays. For the case of the simulated data, the contribution of each
quark flavour is also indicated.   

\begin{figure}
\begin{center}
    \includegraphics[width=0.45\linewidth]{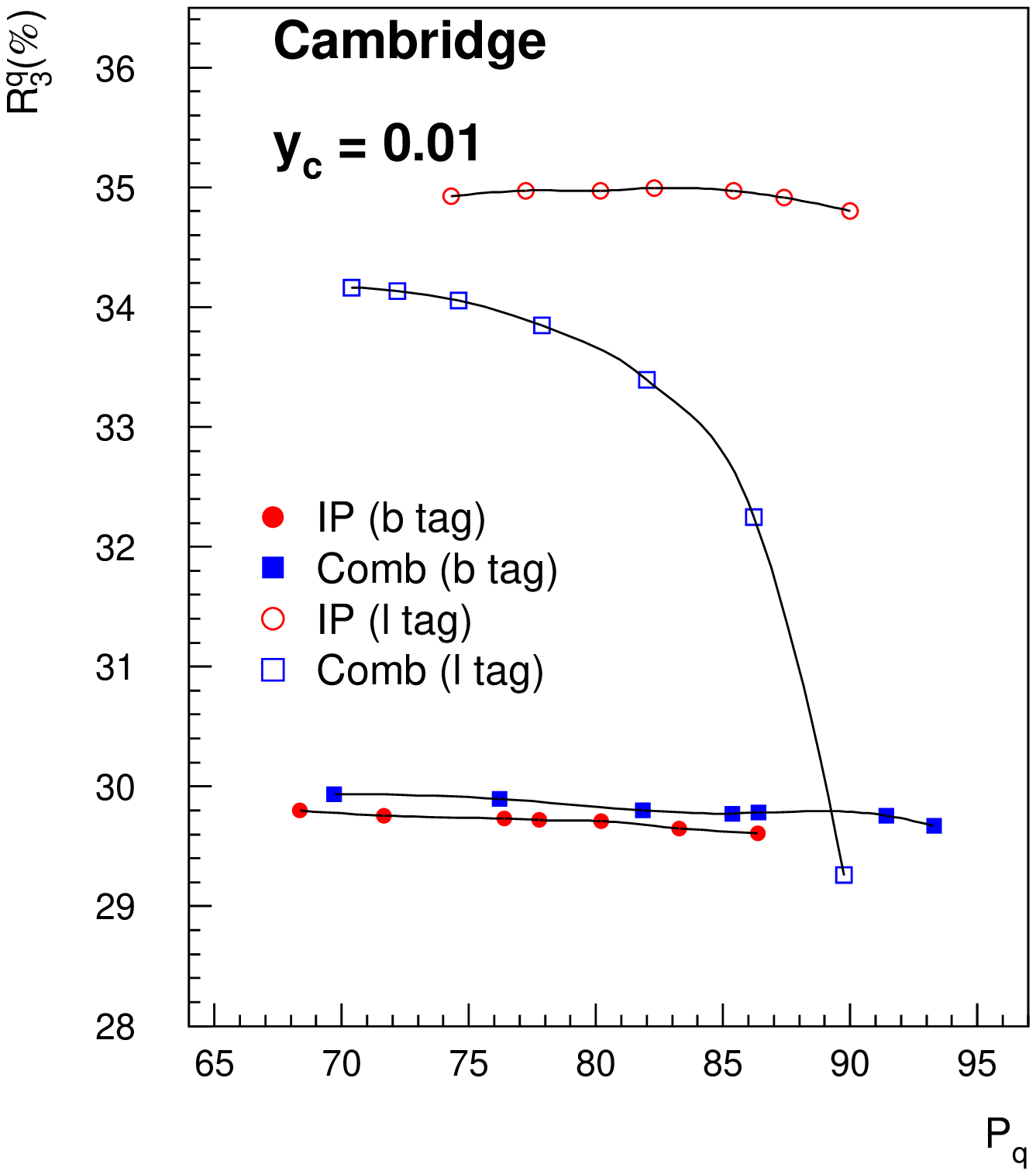} 
    \includegraphics[width=0.45\linewidth]{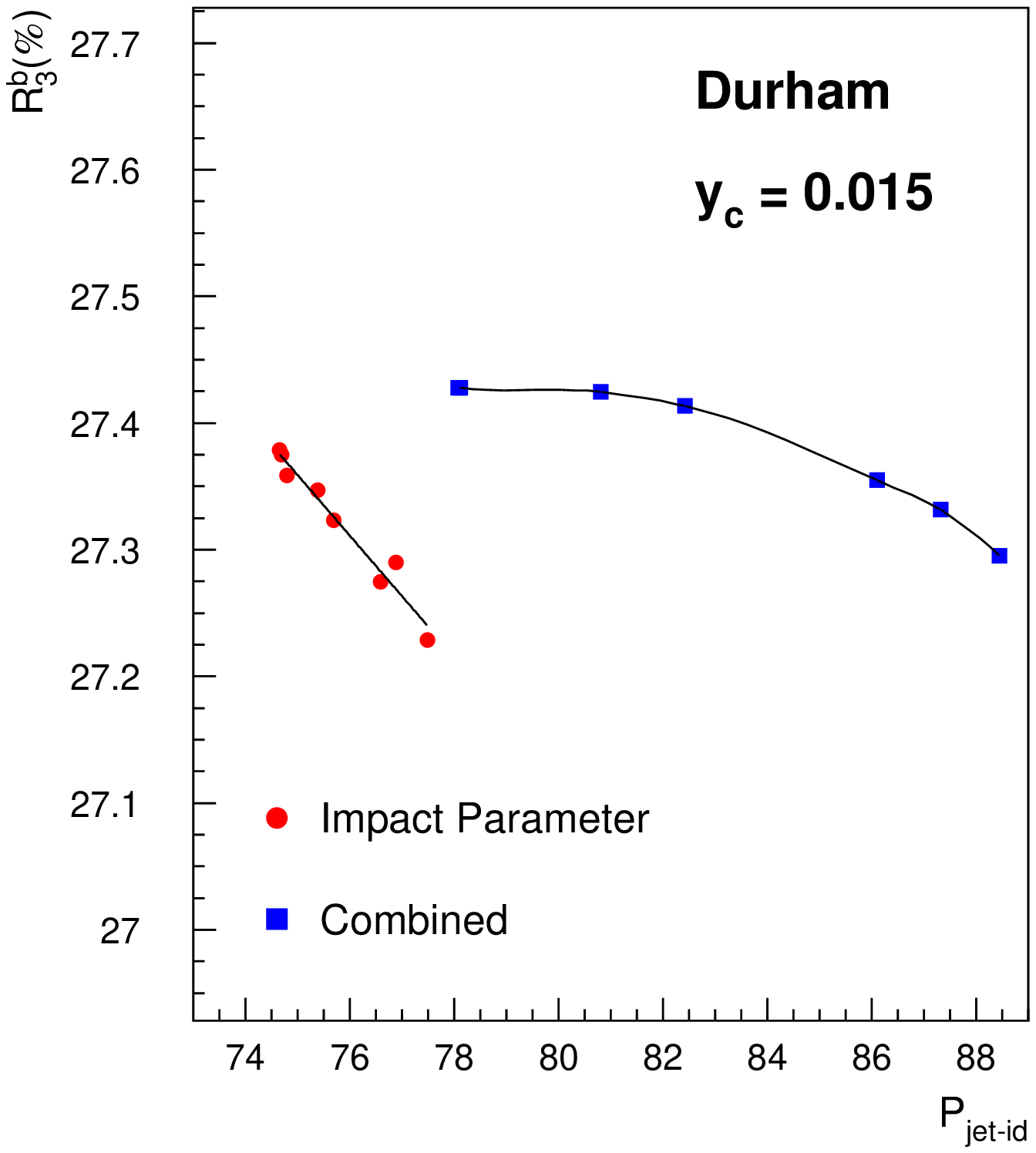}\end{center} 
\caption{(Left) $R_3^q= \Gamma_{3j}^q(y_c) / \Gamma^q$ at parton level as a function of the purity of
the $q$-tagged sample, $P_q$, for $q=b,\ell$, when {\sc Cambridge} is used to
form jets. (Right) $R_3^b$ as a function of the jet identification purity for 
the {\sc Durham} algorithm.}
\label{fig:stability}
\end{figure}

Taking into account the stability of the final result (see Figure
\ref{fig:stability} left), the impact parameter method was used for $\ell$
tagging by imposing $P_E^+>0.07$. The resulting purity of the sample and 
efficiency of the selection were $P_\ell = 82$\% and $\epsilon_\ell$ =
51\%, respectively. For $b$ tagging both techniques were observed to be 
equally stable. The combined method was used requiring $X_{effev} > -0.15$ 
since higher purities could be reached for the same efficiency. 
The final purity of the sample and the total efficiency were 
$P_b =86$\% and $\epsilon_b$ = 47\%, respectively, where this efficiency value
also takes into account the hadronic selection.  

In order to perform the cut on the $b$ quark energies (see Section
3.1), an identification of the gluon and $b$ quark jets was required
for $b$-tagged events. The two tagging techniques can also provide a
discriminant variable per jet and therefore both are available for jet
identification. Again, based on a stability argument (see Figure
\ref{fig:stability} right), the combined technique was used to
identify the pair of jets most likely to come from $b$ quarks by
requiring $X_{effev} > -0.5$ (where now $X_{effev}$ is only computed
with the tracks contained in the pair of jets which gives the maximum
$X_{effev}$). This results in a $b$ jet purity of 81\% per jet in each
event and a tag efficiency of 90\%.

\begin{figure}
\begin{center}
  \includegraphics[width=0.5\linewidth]{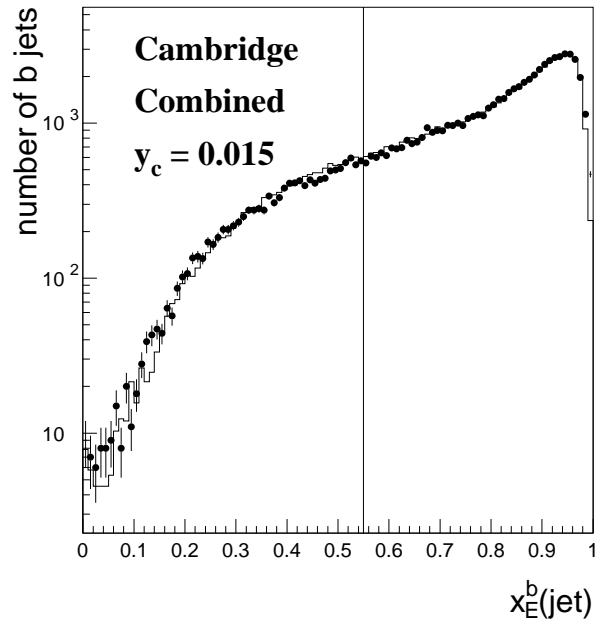} 
\end{center} 
\caption{${\rm x_E^b}(\mathrm{jet})$ distribution for real and 
simulated data for three-jet $b$-tagged events at $y_c=0.015$ for {\sc
Cambridge}.} 
\label{fig:xejet_damc}
\end{figure}

Once the $b$ quark jets were identified their energy was computed from the
jet directions using energy-momentum conservation and assuming massless
kinematics. Figure \ref{fig:xejet_damc} shows the
${\rm x_E^b}(\mathrm{jet})$ distribution for real and simulated data. The cut
${\rm x_E^b}(\mathrm{jet}) \ge 0.55$ was then applied for both $b$ jets. The purity
and contamination factors of the $b$ and $\ell$-tagged samples obtained after
the cut are shown in Table \ref{tab:ciq}. 

\begin{table}[htb]
\begin{center}
\vspace{7mm}
\begin{tabular}{|c|c|c|c|c|}
\hline
Method & Type~$q$~ & ${\ell} \rightarrow q$ (\%) & $c \rightarrow q$ (\%) &
~$b\rightarrow q$  (\%)\\ 
\hline
Impact Parameter & ${\ell}$     & 82 & 15  & 3  \\
Combined & $b$                  &  2 & 7   & 91 \\
\hline
\end{tabular}
\vspace{0.5cm}
\caption{Flavour composition of the samples tagged as
$\ell$ or $b$ quark events. $(\ell,c,b) \rightarrow q$ refers to the fraction of
true $q^\prime$ events in the $q$-tagged sample.}
\label{tab:ciq}
\end{center}
\end{table}

%%%%%%%%%%%%%%%%%%%%%%%%%%%%%%%%%%%%%%%%%%%%%%%%%%%%%%%%%%%%%%%%%%%%%%%%%%%%
\subsection{Data correction}
\label{sec-corr}
%%%%%%%%%%%%%%%%%%%%%%%%%%%%%%%%%%%%%%%%%%%%%%%%%%%%%%%%%%%%%%%%%%%%%%%%%%%%

Once the $b$ and $\ell$ quark hadronic events were selected from the
collected data, the jet-clustering algorithm was applied to get the
$R_3^{b\ell}$ observable at detector level, $R_3^{b\ell-det}$. In
order to bring the observable to parton level the method of the
previous {\sc Delphi} measurement was used \cite{delmbmz}. A
correction to obtain pure $b$ and $\ell$-quark samples was applied in
this procedure and the flavour composition uncertainties were
accounted for by the tagging uncertainty.

The {\sc Delphi} full simulation ({\sc Delsim}), which uses {\sc Jetset
7.3} to generate the events that go through the detector
simulation, was used to compute the detector correction. 
A reweighting of the events was done in order to reproduce the measured
heavy quark gluon-splitting rates \cite{gs_average} 
($g_{c\overline{c}} = 0.0296 \pm 0.0038$ and 
$g_{b\overline{b}} = 0.00254 \pm 0.00051$) in the simulation.

A recent version of {\sc Pythia} 6.131, tuned to 
{\sc Delphi} data \cite{delphi_tuning} and with the
kinematical $b$ quark mass parameter set to $M_b = 4.99 \pm 0.13$ GeV/$c^2$,
was used to get the hadronization correction.   

\begin{figure}
\begin{center}
    \includegraphics[width=0.48\linewidth]{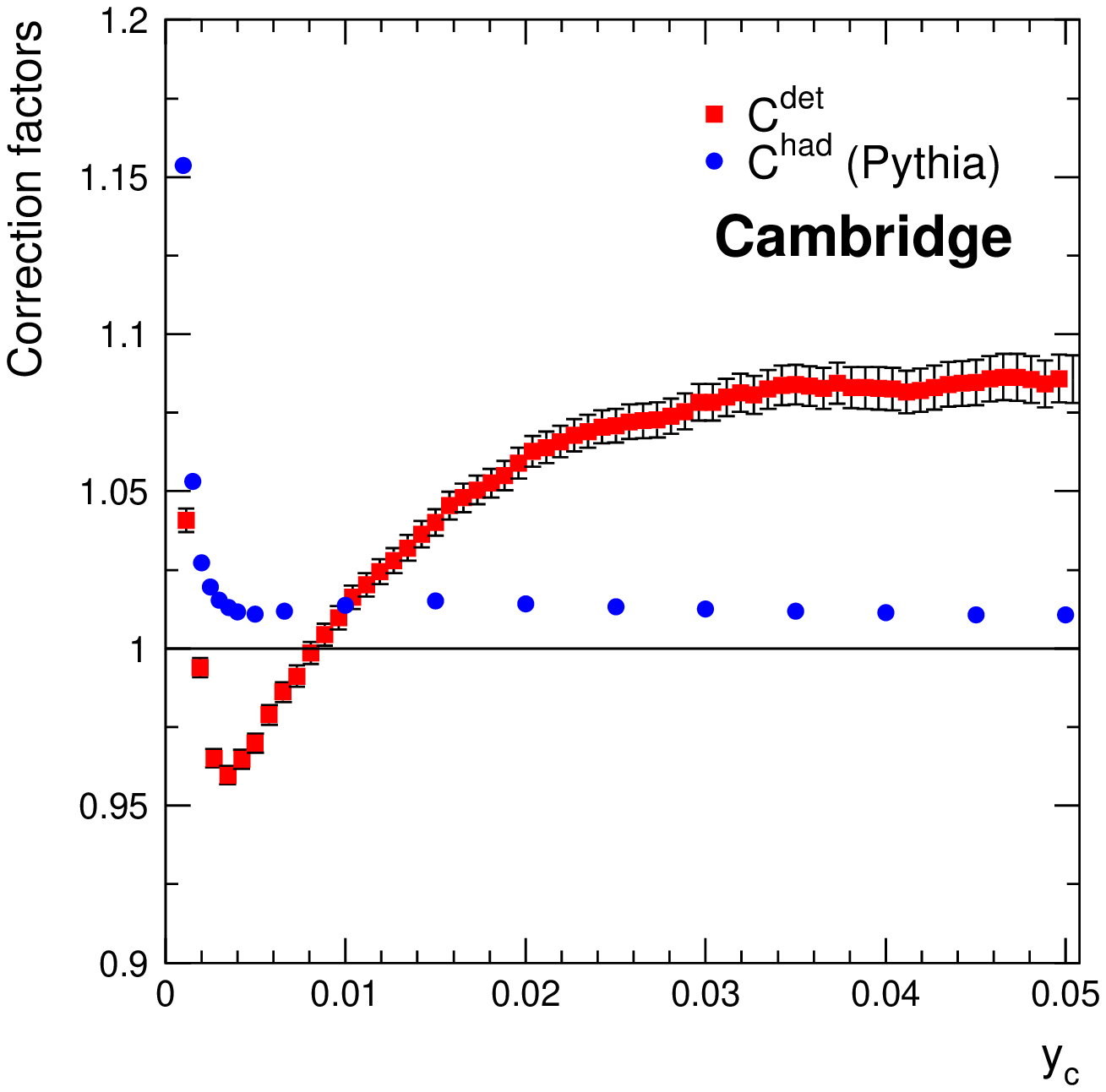}
    \includegraphics[width=0.48\linewidth]{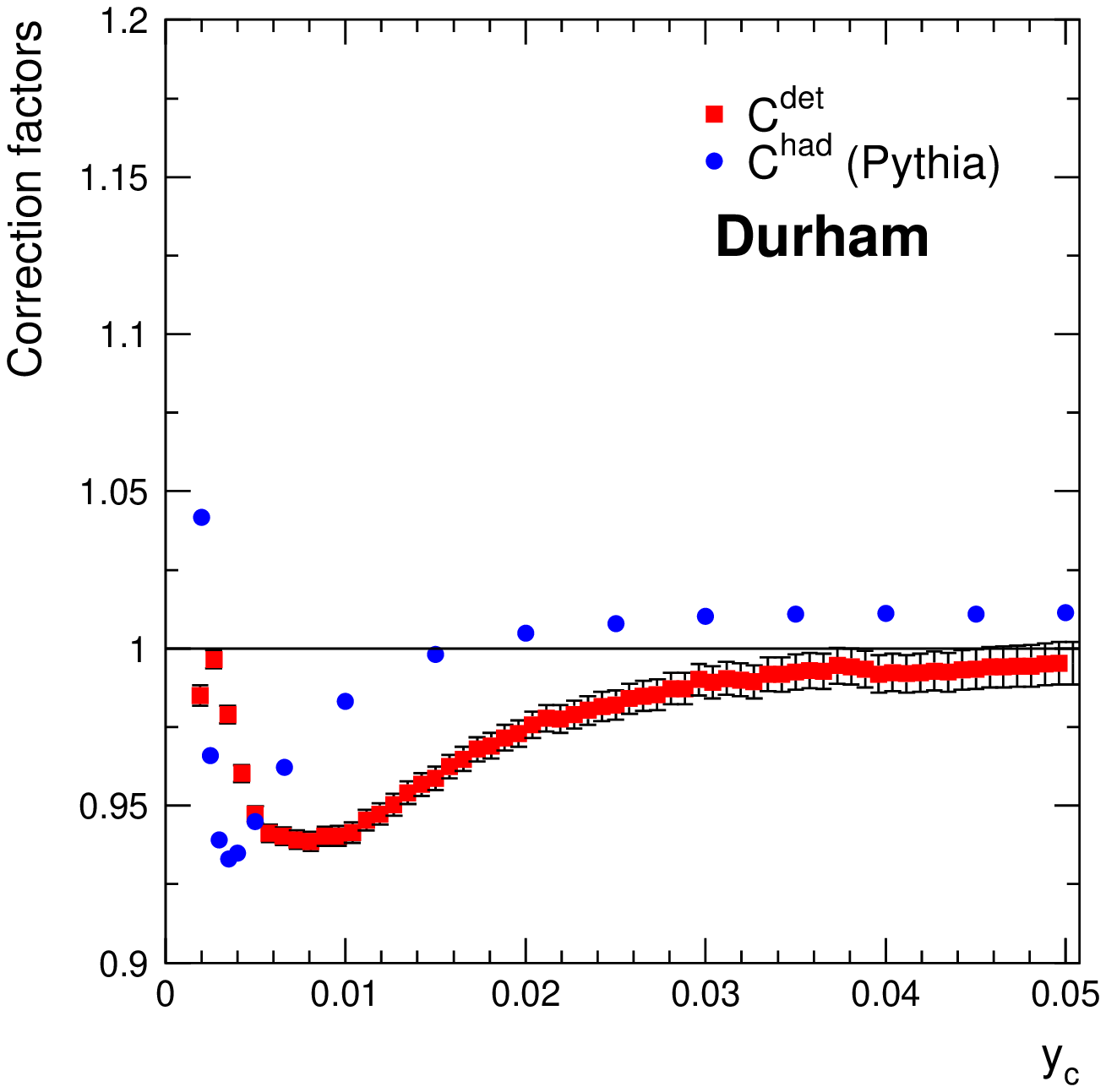}
\end{center} 
\caption{Detector and hadronization corrections applied to the
measured $R_3^{b\ell}$ for {\sc Cambridge} and {\sc Durham}. The
detector correction, $C^{det}$, brings the observable to hadron level,
and the hadronization correction, $C^{had}$, brings it from this stage
to parton level.  }
\label{fig:det_had_corr}
\end{figure}

The magnitude of the detector and hadronization corrections for
$R_3^{b\ell}$ are shown in Figure \ref{fig:det_had_corr}. At the $y_c$
value chosen for the final result ($y_c = 0.0085$ and $y_c = 0.02$ for
{\sc Cambridge} and {\sc Durham}, respectively) the detector
correction is about -2.5\% for {\sc Durham} and 5\permil $~$ for {\sc
Cambridge}. The hadronization correction is 1\% for {\sc Cambridge}
and half as big for {\sc Durham}.

%%%%%%%%%%%%%%%%%%%%%%%%%%%%%%%%%%%%%%%%%%%%%%%%%%%%%%%%%%%%%%%%%%%%%%%%%%%%
\subsection{Experimental uncertainties}
\label{sec-exp_unc}
%%%%%%%%%%%%%%%%%%%%%%%%%%%%%%%%%%%%%%%%%%%%%%%%%%%%%%%%%%%%%%%%%%%%%%%%%%%%

Apart from the statistical uncertainties, different sources of systematic
uncertainties were considered. They can be divided into two groups: those due
to the hadronization correction and those due to the detector correction. 

%\newpage

\begin{itemize}

  \item {\bf hadronization}: 

  The following sources of uncertainties in the hadronization correction have
  been taken into account: 

  \begin{itemize}

    \item uncertainty in the tuned parameters of {\sc Pythia}
    that are relevant in the fragmentation process. This contribution was
    evaluated by varying these parameters ($\Lambda_{\mathrm{QCD}}$, $Q_0$,
    $\sigma_q$, $\epsilon_b$, $a$) within $\pm 1$ standard deviation around their tuned 
    central values, taking into account correlations \cite{delphi_tuning};

    \item uncertainty due to the choice of the hadronization model. It was
    calculated as the standard deviation of the three hadronization
    models used (see Section \ref{sec-model});  

    \item uncertainty from varying the value of the $b$ quark mass parameter 
    in the generator within the error of 0.13 GeV/$c^2$ about its chosen 
    central value of $M_b$ = 4.99 GeV/$c^2$ (see Section \ref{sec-masspythia}).

  \end{itemize}

  \item {\bf detector}: 

  The uncertainties in the detector correction, including selection 
  efficiencies, acceptance effects and the tagging procedure, are due to 
  imperfections in the physics and detector modelling provided in the 
  simulation. The following sources were considered:

  \begin{itemize}

    \item gluon-splitting: The measured $c\overline{c}$ and $b\overline{b}$ 
    gluon-splitting rates were varied within their
    uncertainty and the effect on the measured observable was taken as the
    gluon-splitting error;  

    \item tagging: The related uncertainty was evaluated by varying the 
    tagging and mis-tagging efficiencies within their uncertainties:
    $\Delta \epsilon_b^b/\epsilon_b^b = 3$\% and
    $\Delta \epsilon_b^\ell/\epsilon_b^\ell = \Delta
    \epsilon_b^c/\epsilon_b^c = 8$\% evaluated as in \cite{rb}
    (being $\epsilon_q^{q^{\prime}}$ the fraction of $q^{\prime}$ tagged
    events in the true $q$-quark sample). For this purpose, 
    $\ell$ tagging was considered equivalent to anti-$b$ tagging, i.e. 
    $\Delta \epsilon_\ell^q = \Delta \epsilon_b^q$ for $q=b,c,\ell$ for the 
    same cut value;

    \item jet identification: The cut
    applied to distinguish the $b$ quark jets from the gluon jet in a $b$
    tagged event was varied in order to obtain cut efficiencies (i.e. the 
    fraction of events which pass the cut applied to select the two $b$ quark 
    jets among the 3 jets)  between 80\%
    and almost 100\%. Half of the full variation observed in the measured 
    $R_3^{b\ell}$ at parton level was taken as the uncertainty due to the 
    jet identification. 

  \end{itemize}
\end{itemize}

%%%%%%%%%%%%%%%%%%%%%%%%%%%%%%%%%%%%%%%%%%%%%%%%%%%%%%%%%%%%%%%%%%%%%%%%%%%%
\subsection{Results for $R_3^{b\ell}$ at hadron and parton level}
\label{sec-r3blresults}
%%%%%%%%%%%%%%%%%%%%%%%%%%%%%%%%%%%%%%%%%%%%%%%%%%%%%%%%%%%%%%%%%%%%%%%%%%%%

\begin{figure}
\begin{center}
  \includegraphics[width=0.48\linewidth]{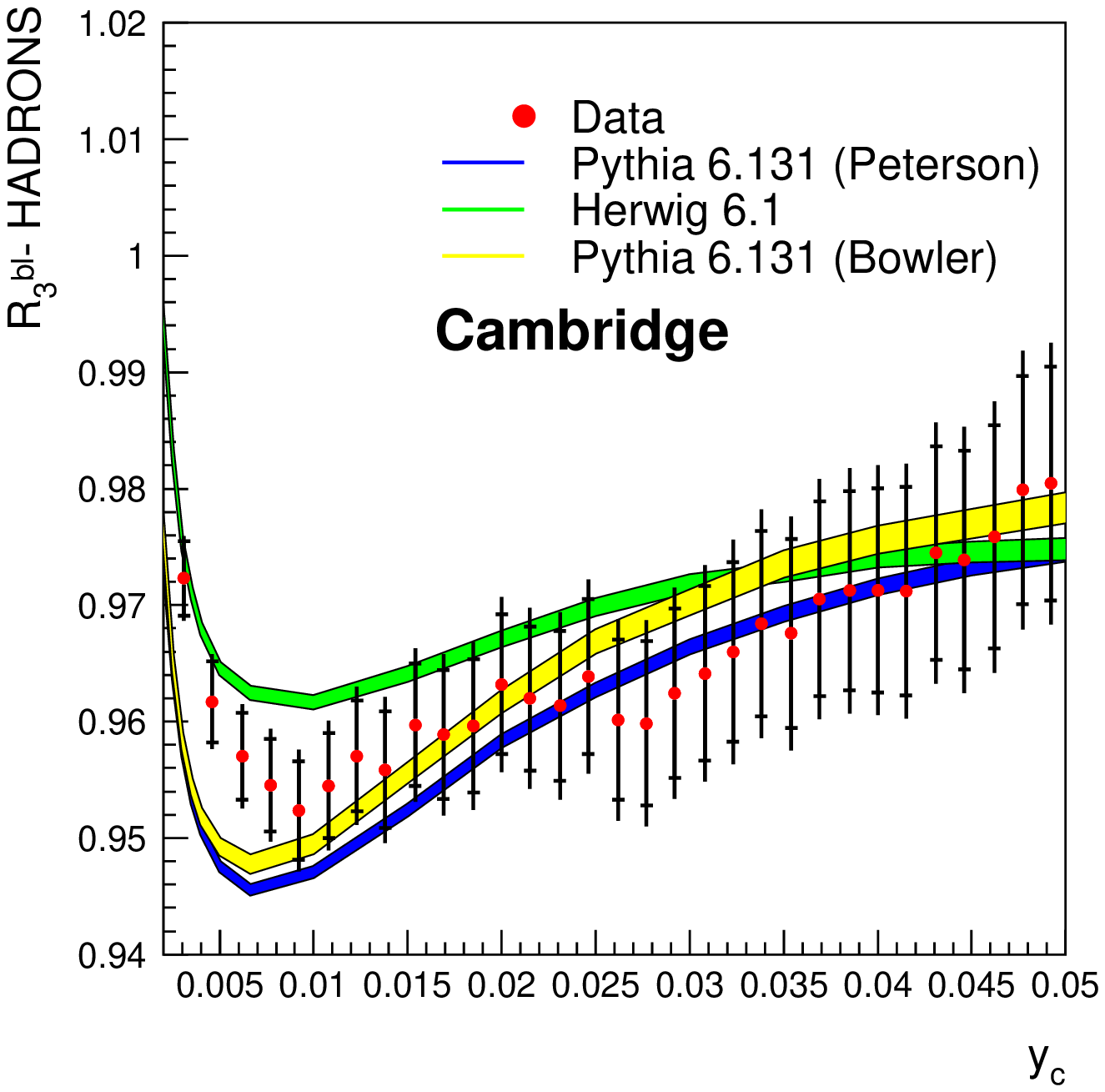}
  \includegraphics[width=0.48\linewidth]{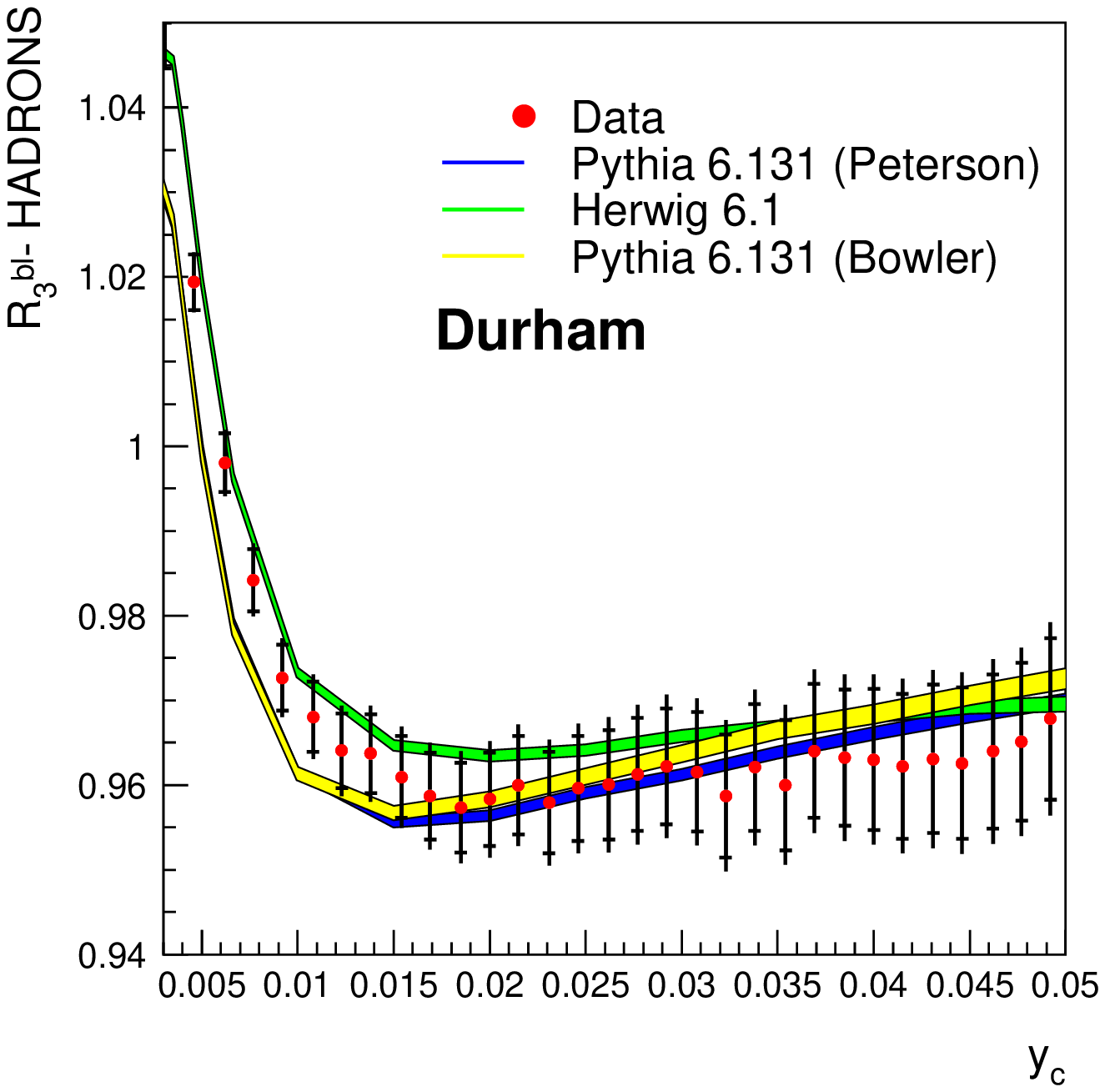}
\end{center} 
\caption{$R_3^{b\ell}$ at hadron level as a function of $y_c$ 
compared with {\sc Pythia} 6.131 (with Peterson and Bowler fragmentation
$b$ functions) and {\sc Herwig} 6.1 predictions, using the {\sc Cambridge} 
(left) and {\sc Durham} (right) jet-clustering algorithms.}
\label{fig:r3blhad}
\end{figure}

Figure \ref{fig:r3blhad} shows, as a function of the $y_c$, the measured
$R_3^{b\ell}$ corrected to hadron level together with the curves predicted by
the {\sc Pythia} and {\sc Herwig} generators. The statistical-only and 
total uncertainties can also be seen in this figure. The {\sc Pythia} curves 
are shown independently for the Peterson and Bowler $b$ fragmentation 
functions. For large values of $y_c$, both generators describe the data well. 
The measured $R_3^{b\ell}$ ratio and its uncertainties are 
also presented in Tables \ref{tabcamhad} and \ref{tabdurhad}.  

% which usually tends to overestimate the suppression of the gluon emission 
%due to the $b$ quark mass in the three jet topology. 

\begin{figure}
\begin{center}
  \includegraphics[width=0.48\linewidth]{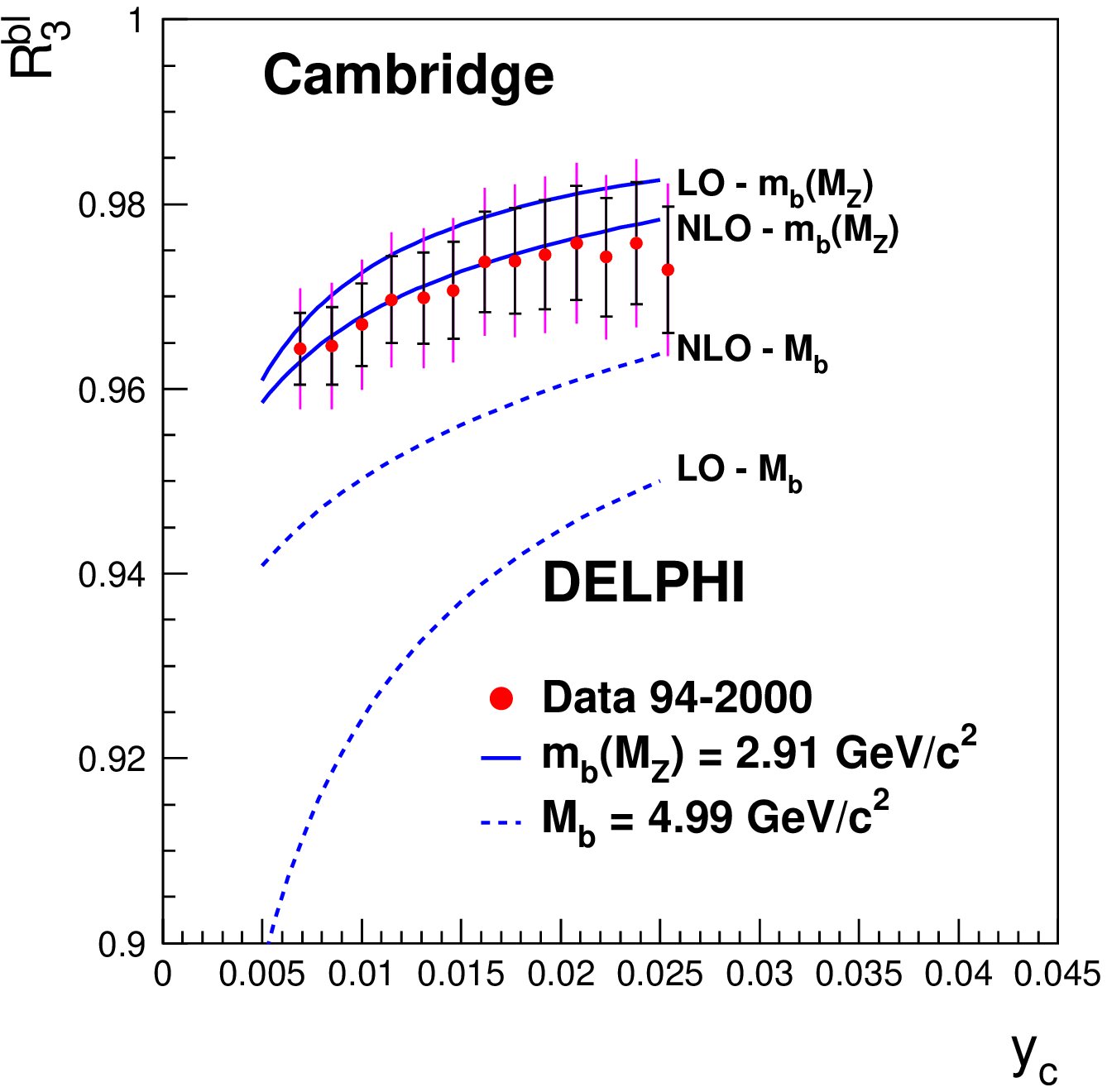}
  \includegraphics[width=0.48\linewidth]{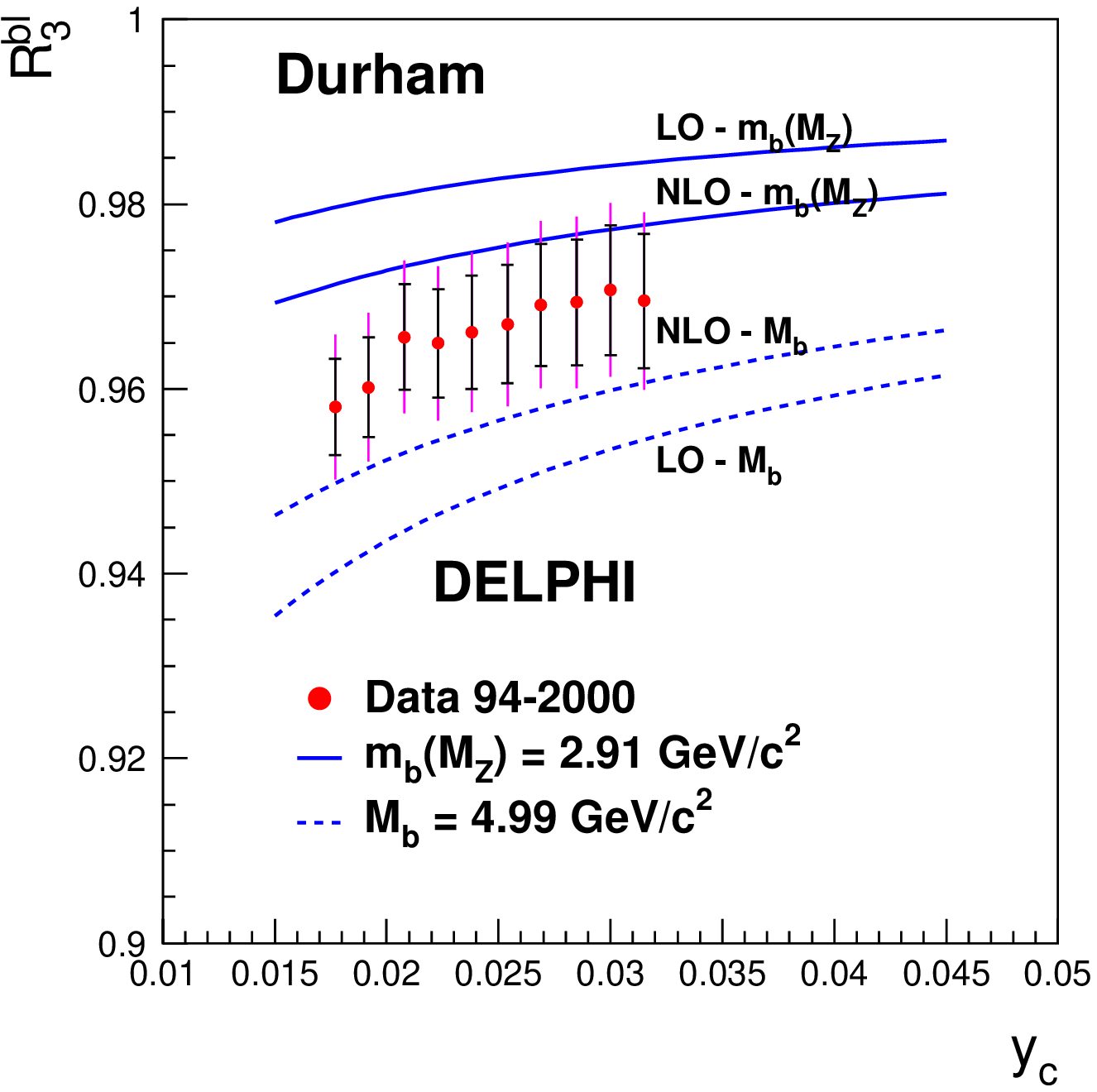}
\end{center} 
\caption{$R_3^{b\ell}$ as a function of $y_c$ obtained at parton level
compared with the LO and NLO theoretical predictions calculated in terms of a
pole mass of $M_b = 4.99$ GeV/$c^2$ and in terms of a running mass of
$m_b(M_Z) = 2.91$ GeV/$c^2$.}
\label{fig:r3blpart}
\end{figure}

\begin{figure}
\begin{center}
  \includegraphics[width=0.48\linewidth]{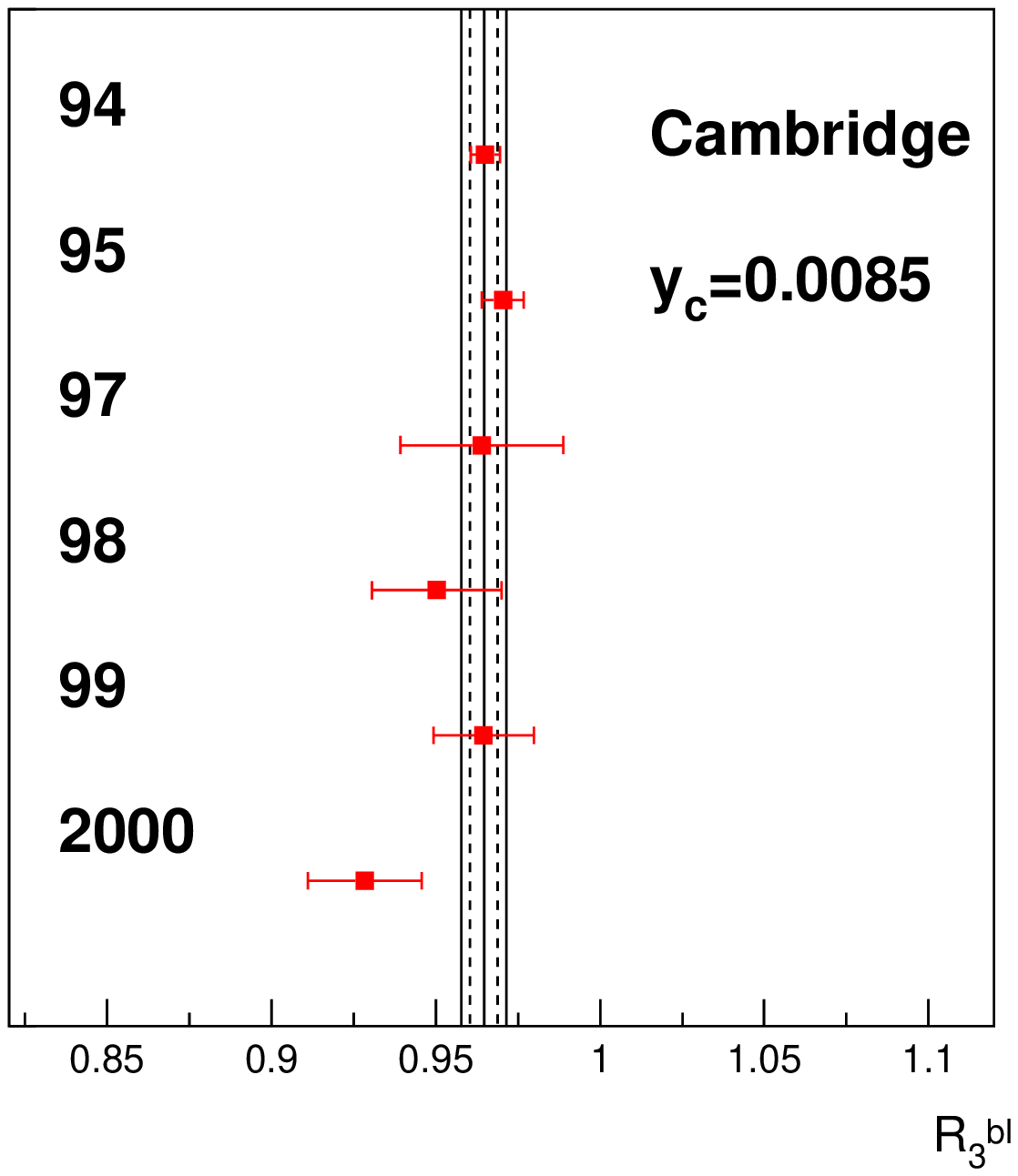}
  \includegraphics[width=0.48\linewidth]{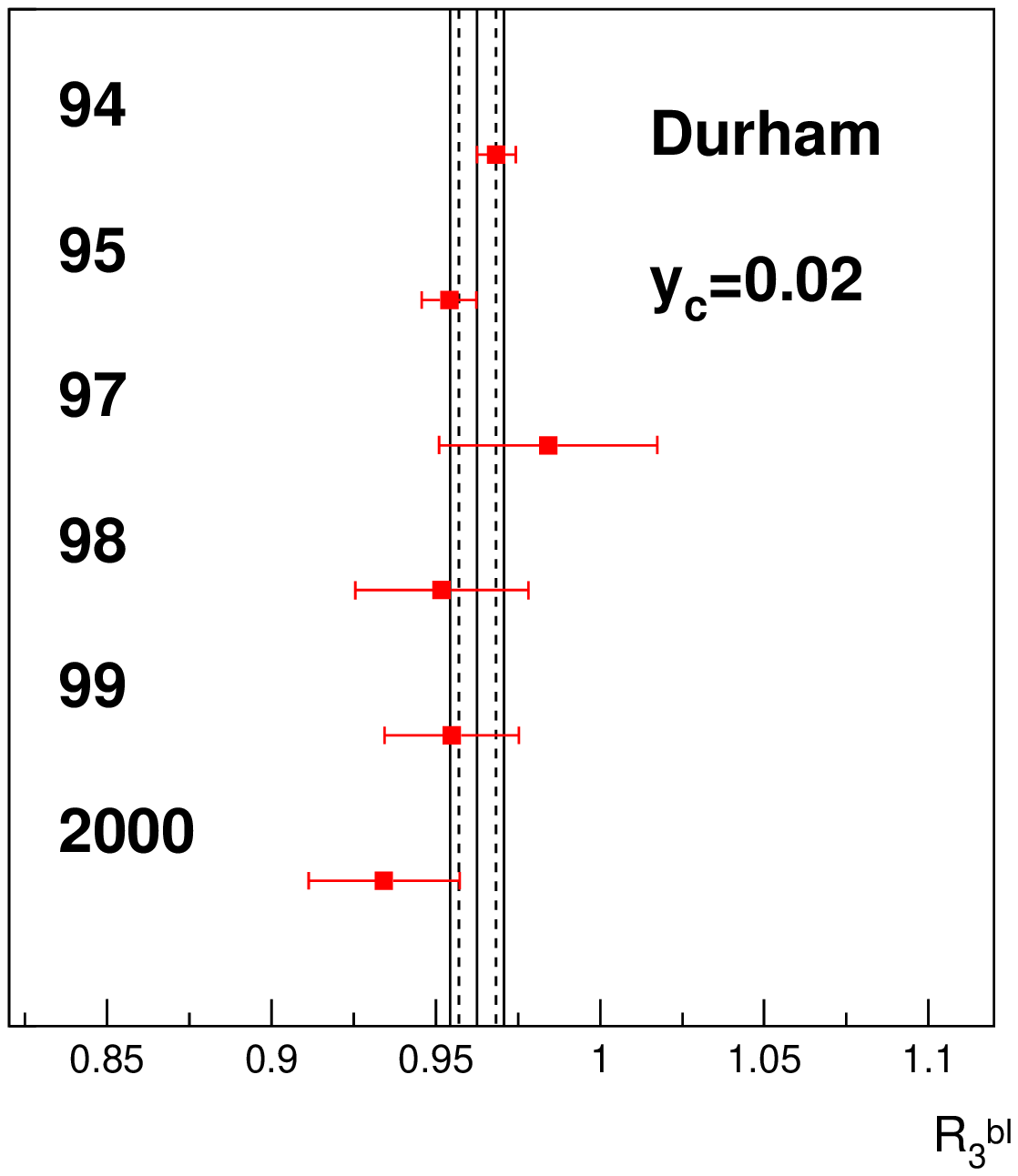}
 \end{center} 
\caption{$R_3^{b\ell}$ at parton level obtained for each analysed year for a
fixed $y_c$ for {\sc Cambridge} (left) and {\sc Durham} (right). The error
bars represent the statistical error. The vertical lines show the average
value with its statistical and total error. The $\chi^2$ per degree of
freedom of the average is 0.7 and 1.2 for {\sc Cambridge} and {\sc Durham}
respectively.} 
\label{fig:r3bl_any}
\end{figure}

The result for $R_3^{b\ell}$ obtained at parton level is shown in
Figure \ref{fig:r3blpart} as a function of $y_c$ together with its
statistical and total uncertainties. The LO and NLO theoretical
predictions in terms of the pole and running masses ($M_b = 4.99$
GeV/$c^2$ and $m_b(M_Z) = 2.91$ GeV/$c^2$) are also shown in the
plot. In the case of {\sc Cambridge} the LO prediction is already
reproducing the measured data, indicating a better convergence in the
theoretical calculations than for {\sc Durham}.  The results for the
individual years of data taking are compatible (see Figure
\ref{fig:r3bl_any}).

\begin{table}[h,t]
\begin{center}
\begin{tabular}{lrrrrr}
\hline
$y_c$ & $R_3^{b\ell-had}$ & $\sigma_{stat}$ & $\sigma_{g-splitting}$ & $\sigma_{tag}$ & $\sigma_{jet-id}$ \\ \hline
0.005 &0.9617  & $\pm$0.0034 & $\pm$0.0003 & $\pm$0.0014 & $\pm$0.0016 \\  
0.01  &0.9544  & $\pm$0.0044 & $\pm$0.0010 & $\pm$0.0025 & $\pm$0.0019 \\ 
0.015 &0.9560  & $\pm$0.0052 & $\pm$0.0014 & $\pm$0.0031 & $\pm$0.0021 \\ 
0.02  &0.9632  & $\pm$0.0059 & $\pm$0.0018 & $\pm$0.0036 & $\pm$0.0024 \\ 
0.025 &0.9639  & $\pm$0.0067 & $\pm$0.0021 & $\pm$0.0039 & $\pm$0.0026 \\  
0.03  &0.9629  & $\pm$0.0074 & $\pm$0.0024 & $\pm$0.0041 & $\pm$0.0029 \\ \hline
\end{tabular}
\caption{$R_3^{b\ell}$ at hadron level at different $y_c$ with jets 
reconstructed with {\sc Cambridge}. }
\label{tabcamhad}
\end{center}
\end{table}

\begin{table}[h,t]
\begin{center}
\begin{tabular}{lrrrrr}
\hline
$y_c$ & $R_3^{b\ell-had}$ & $\sigma_{stat}$ & $\sigma_{g-splitting}$ & $\sigma_{tag}$ & $\sigma_{jet-id}$ \\ \hline
0.005 & 1.0194 & $\pm$0.0033 & $\pm$0.0002 & $\pm$0.0011 & $\pm$0.0008 \\ 
0.01  & 0.9690 & $\pm$0.0039 & $\pm$0.0007 & $\pm$0.0025 & $\pm$0.0010 \\  
0.015 & 0.9613 & $\pm$0.0047 & $\pm$0.0012 & $\pm$0.0031 & $\pm$0.0012 \\ 
0.02  & 0.9583 & $\pm$0.0056 & $\pm$0.0016 & $\pm$0.0036 & $\pm$0.0014 \\ 
0.025 & 0.9596 & $\pm$0.0062 & $\pm$0.0018 & $\pm$0.0040 & $\pm$0.0015 \\ 
0.03  & 0.9611 & $\pm$0.0070 & $\pm$0.0022 & $\pm$0.0043 & $\pm$0.0017\\  
0.035 & 0.9606 & $\pm$0.0076 & $\pm$0.0025 & $\pm$0.0045 & $\pm$0.0019 \\ 
0.04  & 0.9630 & $\pm$0.0083 & $\pm$0.0027 & $\pm$0.0046 & $\pm$0.0021 \\ 
0.045 & 0.9626 & $\pm$0.0089 & $\pm$0.0029 & $\pm$0.0048 & $\pm$0.0022 \\ 
0.05  & 0.9687 & $\pm$0.0097 & $\pm$0.0032 & $\pm$0.0050 & $\pm$0.0024 \\ \hline
\end{tabular}
\caption{$R_3^{b\ell}$ at hadron level at different $y_c$ with jets 
reconstructed with {\sc Durham}. }
\label{tabdurhad}
\end{center}
\end{table}

%%%%%%%%%%%%%%%%%%%%%%%%%%%%%%%%%%%%%%%%%%%%%%%%%%%%%%%%%%%%%%%%%%%%%%%%%%%%
\section{Comparison with NLO massive calculations}
\label{sec-nlo}
%%%%%%%%%%%%%%%%%%%%%%%%%%%%%%%%%%%%%%%%%%%%%%%%%%%%%%%%%%%%%%%%%%%%%%%%%%%%

The measurement of the $R_3^{b\ell}$ observable at parton level obtained in
the previous section, when compared with the NLO massive calculations of
\cite{nlo1,nlo4}, can be used either to extract the $b$ quark mass assuming
$\alpha_s$ universality or to test $\alpha_s$ flavour independence taking the
$b$ quark mass measured at threshold as an input. 

%%%%%%%%%%%%%%%%%%%%%%%%%%%%%%%%%%%%%%%%%%%%%%%%%%%%%%%%%%%%%%%%%%%%%%%%%%%%
\subsection{Determination of the $b$ quark mass}
\label{sec-mb}
%%%%%%%%%%%%%%%%%%%%%%%%%%%%%%%%%%%%%%%%%%%%%%%%%%%%%%%%%%%%%%%%%%%%%%%%%%%%

In order to extract the $b$ quark mass from the experimentally measured 
$R_3^{b\ell}$, a value of $y_c$ must be chosen for both {\sc Cambridge} 
and {\sc Durham} jet algorithms. The value used was that which gave the 
smallest overall uncertainty on the measurement while staying in the region 
where the hadronization correction remains flat. In this way it was also guaranteed to keep far enough from the four-jet region. The selected values found 
to best fulfill 
these requirements were  $y_c = 0.0085$ and $y_c = 0.02$ for {\sc Cambridge} 
and {\sc Durham}, respectively, where the four-jet rates are 4-5$\%$ and 
2-3$\%$ in each case.
 
The $b$ quark pole mass, $M_b$, could be extracted from the measured 
$R_3^{b\ell}$ using the NLO expression of $R_3^{b\ell}$ in terms of $M_b$ 
\cite{nlo1,nlo4}. Theoretical sources of uncertainty were the $\mu$ scale 
dependence, the identification of the $b$ quark mass parameter in the generator
(see Section \ref{sec-masspythia}) and $\alpha_s$.
%The theoretical uncertainty was considered as the contribution of 
%the $\mu$ scale dependence, the $b$ quark mass parameter in the generator 
%and the  $\alpha_s$ uncertainty. 

The measured $b$ quark pole mass was found to be,
\begin{equation}
M_b = 4.19 \pm 0.23~({\rm stat}) \pm 0.17 ~({\rm exp}) 
\pm 0.25 ~({\rm had}) ^{+0.70}_{-0.83}~({\rm theo})~{\rm GeV/c^2} 
\label{eq:pole_camjet}
\end{equation}
when {\sc Cambridge} is used to reconstruct jets with $y_c = 0.0085$ and,

\begin{equation}
M_b = 4.47 \pm 0.31~({\rm stat}) \pm 0.24
~({\rm exp}) \pm 0.24~({\rm had})  ^{+0.64}_{-0.76}~({\rm theo})~{\rm GeV/c^2} 
\label{eq:pole_durham}
\end{equation}
when {\sc Durham} is used instead with $y_c = 0.02$. 
Although compatible within errors, these values are low compared 
with the results obtained when the $b$ pole mass is measured at low energy (as 
for example 4.98 $\pm$ 0.13 GeV/$c^2$ \cite{markus}). The measurement error is 
dominated by the uncertainty from the identification of the $b$ quark mass 
parameter in the generator with the $b$ pole mass which contributes to the 
theoretical error.

The running mass was also obtained using the NLO computations of  
$R_3^{b\ell}$ from references \cite{nlo1,nlo4}, in this case, in terms of 
the running mass at the $M_Z$ energy scale: $m_b(M_Z)$. The theoretical 
uncertainty was estimated by considering the following sources:

\begin{itemize}

  \item {\it dependence on the renormalization scale}: The $\mu$ scale in the 
  theoretical expressions was
  varied from $M_Z/2$ to $2M_Z$ and half of the difference between
  the result obtained on $m_b(M_Z)$ was taken as the $\mu$ scale error;

  \item {\it mass ambiguity}: Starting from the NLO calculation of 
  $R_3^{b\ell}$ in terms of the pole mass $M_b$, the value of $M_b$ could be 
  extracted and transformed to $m_b(M_b)$ which was later evolved to 
  $m_b(M_Z)$ by means of the Renormalization Group Equations.
  This is also a valid procedure to extract $m_b(M_Z)$. At infinite orders 
  the result derived in this way and that obtained directly from the original 
  NLO calculation in terms of the running mass 
  should be the same. The difference between the results obtained 
  from the two procedures was then also considered as
  a conservative indication of the size of the unknown higher order 
  corrections;

  \item {\it $\alpha_s$}: $\alpha_s(M_Z) = 0.1183 \pm 0.0027$
  \cite{alphas} was varied within its uncertainty. The spread of 
  values obtained for $m_b(M_Z)$ was considered as the error due to the 
  $\alpha_s$ uncertainty. 

\end{itemize}

%\begin{figure}
%\begin{center}
%  \includegraphics[width=0.48\linewidth]{exp_errors_camjet}
%  \includegraphics[width=0.48\linewidth]{exp_errors_durham}
%\end{center} 
%\caption{Total, statistical and systematic uncertainty in the 
%measurement of $m_b(M_Z)$ as a function of $y_c$ using {\sc Cambridge} (left) 
%and {\sc Durham} (right) jet clustering algorithms. The vertical line 
%represents the $y_c$ value chosen to give the final $m_b(M_Z)$.}
%\label{fig:mb_errors}
%\end{figure} 

The results obtained for $m_b(M_Z)$ were,
\begin{equation}
m_b(M_Z) = 
2.96 \pm 0.18~({\rm stat}) \pm 0.13 ~({\rm exp})
\pm 0.19 ~({\rm had}) ^{+0.04}_{-0.22} ~({\rm theo})~{\rm GeV}/c^2 
\label{eq:mb_camjet}
\end{equation}
when {\sc Cambridge} was used to reconstruct jets and,
\begin{equation}
m_b(M_Z) = 
3.42 \pm 0.25~({\rm stat}) \pm 0.18 ~({\rm exp})
\pm 0.20 ~({\rm had})  ^{+0.10}_{-0.45} ~({\rm theo})~{\rm GeV}/c^2 
\label{eq:mb_durham}
\end{equation}
in the case {\sc Durham} was the algorithm employed. 

The theoretical uncertainty expressed in this way is highly asymmetric due to 
the mass ambiguity. Hence the interval covered by the extreme values of the 
theoretical uncertainty originating from this mass ambiguity was considered 
as the whole range of theoretical 
uncertainty and the measurement of $m_b(M_Z)$ was set to the mean value of this
region. The effect on the mass value is a shift in the order of 
$\sim$ --100 (--200) MeV/$c^2$ 
for {\sc Cambridge} ({\sc Durham}). The same criteria were also adopted in 
previous work \cite{delmbmz,alephmb} and in the present case leads to:
\begin{equation}
m_b(M_Z) = 
2.85 \pm 0.18~({\rm stat}) \pm 0.13 ~({\rm exp})
\pm 0.19 ~({\rm had}) \pm 0.12 ~({\rm theo})~{\rm GeV}/c^2 
\label{eq:mb_camjet2}
\end{equation}
when {\sc Cambridge} was used to reconstruct jets and,
\begin{equation}
m_b(M_Z) = 
3.20 \pm 0.25~({\rm stat}) \pm 0.18 ~({\rm exp})
\pm 0.20 ~({\rm had}) \pm 0.24 ~({\rm theo})~{\rm GeV}/c^2 
\label{eq:mb_durham2}
\end{equation}
if the {\sc Durham} algorithm was used. 

The contribution of the individual uncertainties is given in Table
\ref{tabdurcam}. The result obtained with {\sc Cambridge} is more precise than
the one obtained with {\sc Durham} mainly because of the smaller
theoretical uncertainty, leading to a total error of $\pm 0.32$ GeV/$c^2$
instead of $\pm 0.44$ GeV/$c^2$.

\begin{table}[h,t]
\begin{center}
\begin{tabular}{lrrr}
\hline
{\sc Cambridge} & $R_3^{b\ell-had}$ & $R_3^{b\ell-part}$ & $m_b(M_Z)$   \\
                & $(y_c=0.0085)$    & $(y_c=0.0085)$     & GeV/$c^2$    \\
\hline
 & & \\ 
Value                  & 0.9527      & 0.9646         &  2.85             \\
 & & \\ \hline 
Statistical Data       & $\pm0.0033$ & $\pm0.0034$    & $\pm$0.14  \\
Statistical Simulation & $\pm0.0024$ & $\pm0.0025$    & $\pm$0.11 \\ \hline
Total statistical      & $\pm0.0041$ & $\pm0.0042$    & $\pm$0.18 \\ \hline 
Fragmentation Tuning   & --~~~       & $\pm0.0010$    & $\pm$0.04 \\
Fragmentation Model    & --~~~       & $\pm0.0025$    & $\pm$0.11 \\
Mass parameter         & --~~~       & $\pm0.0036$    & $\pm$0.16 \\ \hline
Total hadronization    & --~~~       & $\pm0.0045$    & $\pm$0.19 \\ \hline
Gluon-Splitting        & $\pm$0.0008 & $\pm0.0008$    & $\pm$0.03 \\
Tagging                & $\pm$0.0022 & $\pm0.0021$    & $\pm$0.09 \\
Jet identification     & $\pm0.0018$ & $\pm0.0020$    & $\pm$0.09\\ \hline
Total experimental     & $\pm0.0030$ & $\pm0.0030$    & $\pm0.13$ \\ \hline
Mass Ambiguity         &   --~~~ &   --~~~            & $\pm0.11$ \\
$\mu$-scale ($0.5 \leq \mu/M_Z \leq 2$) &   --~~~ &   --~~~ & $\pm0.04$ \\
$\alpha_s(M_Z)$        &   --~~~ &   --~~~            & $\pm$0.01 \\ \hline
Total theoretical      &   --~~~ &   --~~~            & $\pm0.12$ \\ \hline
\hline
\hline
{\sc Durham}  & $R_3^{b\ell-had}$ & $R_3^{b\ell-part}$ & $m_b(M_Z)$  \\
              & $(y_c=0.02)$      & $(y_c=0.02)$       & GeV/$c^2$   \\
\hline
 & & \\ 
Value                  & 0.9583  & 0.9626          &  3.20              \\
 & & \\ \hline
Statistical Data       & $\pm0.0045$ & $\pm0.0045$ & $\pm$0.20 \\
Statistical Simulation & $\pm0.0033$ & $\pm0.0033$ & $\pm0.15$ \\ \hline
Total statistical      & $\pm0.0056$ & $\pm0.0056$ & $\pm$0.25\\ \hline 
Fragmentation Tuning   & --~~~       & $\pm0.0015$ & $\pm$0.07\\
Fragmentation Model    & --~~~       & $\pm0.0022$ & $\pm$0.10\\
Mass parameter         & --~~~       & $\pm0.0034$ & $\pm$0.15 \\ \hline
Total hadronization    & --~~~       & $\pm0.0042$ & $\pm$0.20 \\ \hline
Gluon-Splitting        & $\pm0.0016$ & $\pm0.0016$ & $\pm$0.07\\
Tagging                & $\pm$0.0036 & $\pm$0.0035 & $\pm$0.15 \\
Jet identification     & $\pm0.0014$ & $\pm0.0018$ & $\pm$0.08\\ \hline
Total experimental     & $\pm$0.0041 & $\pm0.0042$ & $\pm$0.18 \\ \hline
Mass Ambiguity         &   --~~~ &   --~~~                & $\pm$0.22 \\
$\mu$-scale ($0.5 \leq \mu/M_Z \leq 2$)&   --~~~ &  --~~~~& $\pm$0.10 \\
$\alpha_s(M_Z)$        &   --~~~ &   --~~~                & $\pm$0.02 \\ \hline
Total theoretical      &   --~~~ &   --~~~                & $\pm0.24$ \\ \hline
\hline
\end{tabular}
\caption{Values of $R_3^{b\ell}$ at hadron and parton level and of $m_b(M_Z)$ 
obtained with {\sc Cambridge} and {\sc Durham} algorithms and the break-down 
of their associated errors (statistical and systematic) for $y_c=0.0085$ and $y_c=0.02$ respectively.}
\label{tabdurcam}
\end{center}
\end{table}

%******************************************************************************
\subsection{Test of $\alpha_s$ flavour independence}
\label{sec-alpha}
%******************************************************************************

The measurement of $R_3^{b\ell}$ can alternatively be used to test $\alpha_s$
flavour independence exploiting the relation introduced in \cite{delmbmz}:

\begin{equation}
\alpha_s^b/\alpha_s^\ell = R_3^{b\ell} - H(m_b(M_Z)) + A \cdot 
\frac{\alpha_s(M_Z)}{\pi} ( R_3^{b\ell} - H(m_b(M_Z)) -1 ),
\end{equation}
where $H(m_b(M_Z))$ is the theoretical mass correction and the factor $A$ 
depends on the jet reconstruction algorithm and $y_c$, taking values between 
2 and 6 for all possible circumstances of the present analysis. 

Taking the average $b$ quark mass from low energy measurements, 
$m_b(m_b) = 4.24 \pm 0.11$ GeV/$c^2$ \cite{bmass_02}, as the input $b$ mass value, the 
ratio $\alpha_s^b/\alpha_s^\ell$ is found to be,
\begin{equation}
\alpha_s^b/\alpha_s^\ell = 0.999 \pm 0.004~({\rm stat}) \pm
0.005~({\rm syst}) \pm 0.003~({\rm theo})
\end{equation}
for {\sc Cambridge} and
\begin{equation}
\alpha_s^b/\alpha_s^\ell = 0.990 \pm 0.006 ~({\rm stat}) \pm
0.006~({\rm syst}) \pm 0.005 ~({\rm theo})
\end{equation}
for {\sc Durham}. These results verify $\alpha_s$ universality at a
precision level of 7-9\permil.

%******************************************************************************
%\subsection{The anomalous mass dimension: the $\gamma_0$-function}
%\label{sec-gamma}
%******************************************************************************

%Considering the well known one-loop equation for the energy evolution of the 
%$b$ running mass in the $\overline{\rm MS}$ renormalization scheme:

%
%\begin{equation}
%m(Q)=m(Q_0)K(Q)^{-2\gamma_0/\beta_0}
%\end{equation}
%
%with $K(Q)=\alpha_s(Q_0) / \alpha_s(Q)$ and the anomalous dimension functions
%defined as $\beta_0=11-2N_F/3$ ($N_F$= number of active flavours) and $\gamma_0$
%a value of:

%\begin{equation}
%\gamma_0=2.3 \pm 0.6
%\end{equation}
%
%can be extracted from the above results which agrees with the theoretical 
%expectation of Quantum Chromodynamics of $\gamma_0=2$.

%It should be noticed that the experimental observables used 
%to extract this value of $\gamma_0$ are different: $R_3^{b\ell}$ at high energy 
%and the spectra of hadronic bound states or the moments of the spectrum of the 
%$B$ decay products, at low energy. A more direct test of this parameter could be 
%provided measuring the same observable at different energy scales as it done in 
%the case of the $\beta$ function \cite{delphi_beta} using the thrust 
%distribution. This however 

%\newpage
%%%%%%%%%%%%%%%%%%%%%%%%%%%%%%%%%%%%%%%%%%%%%%%%%%%%%%%%%%%%%%%%%%%%%%%%%%%
\section{Conclusions and discussion}
\label{sec-conclu}
%%%%%%%%%%%%%%%%%%%%%%%%%%%%%%%%%%%%%%%%%%%%%%%%%%%%%%%%%%%%%%%%%%%%%%%%%%%

A new determination of the $b$ quark mass at the $M_Z$ scale has been
performed with the {\sc Delphi} detector at LEP. 
%The data used in this analysis
%only include part of those employed in the previous DELPHI measurement 
%\cite{delmbmz} corresponding to those collected during 1994 which made use
%of the doubled-sided vertex detectors. 
The same observable as for the previous {\sc Delphi} measurement \cite{delmbmz}
was studied, now also using the {\sc Cambridge} jet 
clustering algorithm in addition to {\sc Durham}.
The results obtained with {\sc Cambridge} for $m_b(M_Z)$ were found to be 
more precise, giving:

\begin{equation}
m_b(M_Z)= 2.85 \pm 0.32~{\rm GeV}/c^2.
\end{equation}
This constitutes a substantial improvement with respect to the previous 
{\sc Delphi} measurement in which $m_b(M_Z)$ was determined to be 
$2.67 \pm 0.50$ GeV/$c^2$. This is mainly due to the improved evaluation 
of systematic errors as has been described in this paper.

When using the theoretical prediction of  $R_3^{b\ell}$ for the 
{\sc Cambridge} algorithm the data are reasonably well described by the 
theoretical calculation, already at leading order, using the value 
$m_b(M_Z)$ = 2.91 ${\rm GeV}/c^2$ inferred from the low energy 
measurements (see Figure \ref{fig:r3blpart}). The higher-order terms 
contributing to the calculation of the observable appear to already be 
accounted for in the running of the mass and therefore a faster convergence 
seems to be achieved in comparison with the 
$b$ pole mass. However for {\sc Durham} the situation and the behaviour 
are different as in fact both theoretical predictions at LO are equally 
distant from the data using both mass definitions and NLO calculations 
are certainly needed to describe the data. The value for the $b$ 
pole mass determined in this case was:
\begin{equation}
M_b = 4.19 ^{+0.79}_{-0.91}~{\rm GeV}/c^2. 
\end{equation}
%

%This result is less precise than the one obtained for the $b$ running 
%mass. The values obtained for the pole mass are low but still 
%compatible within errors with the theoretical expectations and the 
%measurements from the low energy data.
 
The present measurement has been performed in a restricted region of the phase
space to have a better control of the fragmentation process. 
%The statistics were increased by analysing data collected at the $Z$ peak
%from 1994 until 2000.
New versions of the generators, {\sc Pythia} 6.131 
and {\sc Herwig} 6.1, where mass effects are much better reproduced, have 
been used to correct the data.

The study of the way mass effects are implemented in the generators, described 
in Section \ref{sec-masspythia}, has led to a more reliable hadronization 
correction. The pole mass definition was shown to be the one to be used 
in the generator and the uncertainty of this identification on the present 
analysis has been quantified. It constitutes the dominant source 
of the present error. 
The effect of the $b\overline{b}$ and $c\overline{c}$ gluon-splitting
rate uncertainties of the Monte Carlo on the detector correction has 
also been taken into account. The observable $R_3^{b\ell}$ is also presented 
at hadron level for different $y_c$ values in view of future versions of the generators
with a better understanding of the hadronization process which could 
then allow for an improved measurement.

\begin{figure}
\begin{center}
  \includegraphics[width=0.9\linewidth]{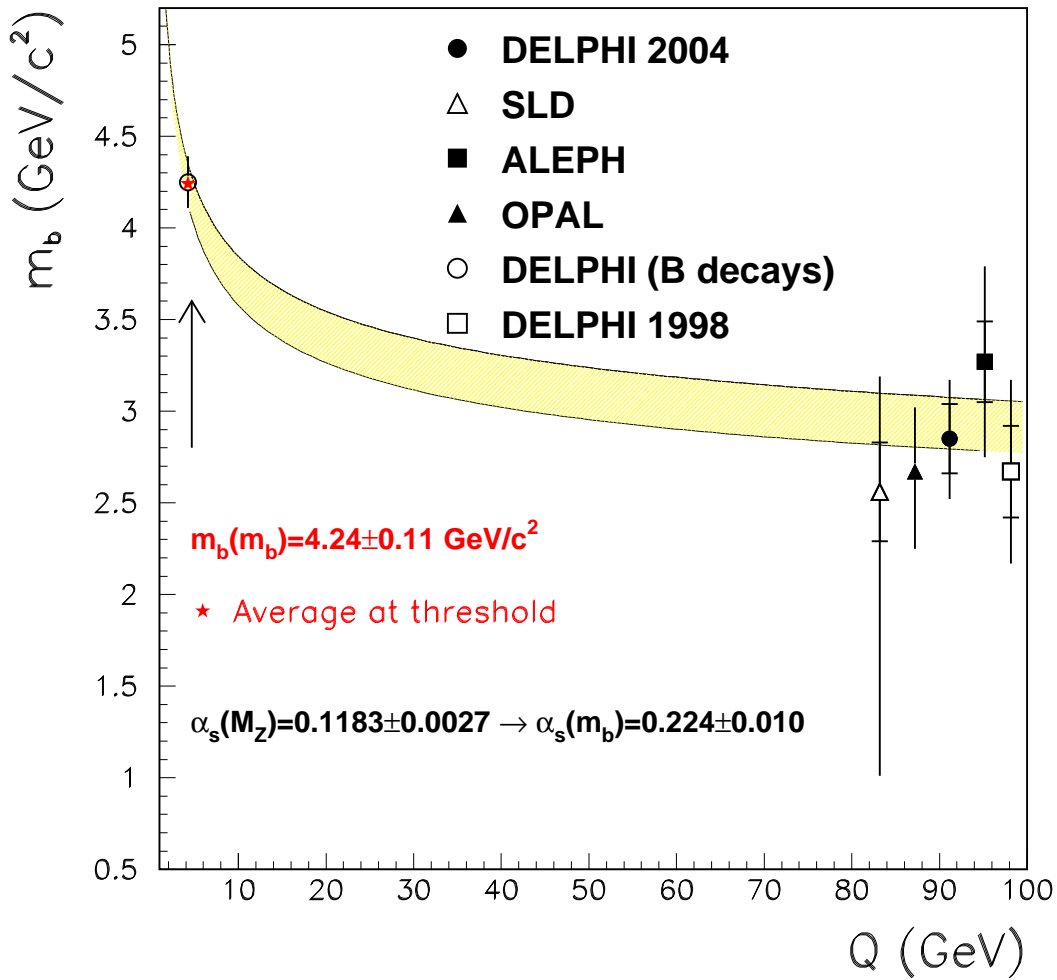}
\end{center} 
\caption{The evolution of $m_b(Q)$ as a function of the energy scale $Q$. 
The $m_b(M_Z)$ measured by LEP and SLC are displayed together with their 
total and statistical errors. The shaded area corresponds to the band 
associated to $m_b(Q)$ when evolving the average value obtained at $m_b(m_b)$ 
\cite{bmass_02} up to the $M_Z$ scale using the QCD Renormalization Group 
Equations with $\alpha_s(M_Z)= 0.1183\pm 0.0027$. All these measurements are 
performed at the $M_Z$ energy scale but for display reasons they are plotted at 
different scales. The result obtained using 
{\sc Delphi} data at low energy from semileptonic $B$ decays 
\cite{lambda_delphi} is also shown.}
\label{fig:mbmz}
\end{figure}

The result obtained by this analysis with {\sc Cambridge} for $m_b(M_Z)$ is 
shown in Figure \ref{fig:mbmz}, together with other LEP and SLC
determinations at the $M_Z$ scale. It is compatible with the other measurements
and is the most precise. The data collected by {\sc Delphi} have 
also been used to determine the $b$ quark mass at a lower energy scale near 
threshold using semileptonic $B$ decays \cite{lambda_delphi}. 
The value obtained in that analysis is also shown. 
The difference between the two measurements is significantly larger than 
the overall uncertainty:

\begin{equation}
\Delta(m_b(m_b)-m_b(M_Z)) = 1.41 \pm 0.36 ~{\rm GeV}/c^2.
\end{equation}
Hence, for the first time, the same experimental data allow values
for the $b$ quark mass to be extracted at two different energy scales. The results obtained
agree with the QCD expectation when using the Renormalization Group Equation 
predictions at the two relevant energy scales of the processes involved. These
observations together with the  average  value of the $b$ quark mass determinations
at threshold \cite{bmass_02}, $m_b(m_b)$, are shown at their corresponding scales 
in Figure \ref{fig:mbmz}. 

Alternatively, universality of the strong 
coupling constant has also been verified with a precision of 7\permil.

For data combination purposes, the above results supersede the previous DELPHI
measurements on this subject \cite{delmbmz}.

\clearpage
\newpage

%         Modified on 04-06-1999 by dimartino
%-------------------------------------------------------------------
\subsection*{Acknowledgements}
\vskip 3 mm
 We are greatly indebted to our technical 
collaborators, to the members of the CERN-SL Division for the excellent 
performance of the LEP collider, and to the funding agencies for their
support in building and operating the DELPHI detector.\\
We acknowledge in particular the support of \\
Austrian Federal Ministry of Education, Science and Culture,
GZ 616.364/2-III/2a/98, \\
FNRS--FWO, Flanders Institute to encourage scientific and technological 
research in the industry (IWT), Belgium,  \\
FINEP, CNPq, CAPES, FUJB and FAPERJ, Brazil, \\
Czech Ministry of Industry and Trade, GA CR 202/99/1362,\\
Commission of the European Communities (DG XII), \\
Direction des Sciences de la Mati$\grave{\mbox{\rm e}}$re, CEA, France, \\
Bundesministerium f$\ddot{\mbox{\rm u}}$r Bildung, Wissenschaft, Forschung 
und Technologie, Germany,\\
General Secretariat for Research and Technology, Greece, \\
National Science Foundation (NWO) and Foundation for Research on Matter (FOM),
The Netherlands, \\
Norwegian Research Council,  \\
State Committee for Scientific Research, Poland, SPUB-M/CERN/PO3/DZ296/2000,
SPUB-M/CERN/PO3/DZ297/2000, 2P03B 104 19 and 2P03B 69 23(2002-2004)\\
FCT - Funda\c{c}\~ao para a Ci\^encia e Tecnologia, Portugal, \\
Vedecka grantova agentura MS SR, Slovakia, Nr. 95/5195/134, \\
Ministry of Science and Technology of the Republic of Slovenia, \\
CICYT, Spain, AEN99-0950 and AEN99-0761,  \\
The Swedish Research Council,      \\
Particle Physics and Astronomy Research Council, UK, \\
Department of Energy, USA, DE-FG02-01ER41155. \\
EEC RTN contract HPRN-CT-00292-2002. \\

%=========================================================================%

We are specially grateful to A. Santamar\'{\i}a and G. Rodrigo for providing
the NLO massive calculations that made this measurement possible. 
We are also indebted to T. Sj\"{o}strand for his help in understanding
how mass effects are implemented in {\sc Pythia}. 
We would also like to thank G. Dissertori for the continuous feedback and
J. Portoles and M. Eidem\"{u}ller for their information about the $b$ pole
mass.

%=========================================================================%
\end{document}